\documentclass[
reprint,
superscriptaddress,
prb,
amsmath,amssymb,
longbibliography
]{revtex4-2}
\usepackage{comment}
\usepackage{graphicx}
\usepackage{dcolumn}
\usepackage{bm}
\usepackage{xspace}
\usepackage{hyperref}
\usepackage{xcolor}
\usepackage[utf8]{inputenc}
\usepackage[T1]{fontenc}
\usepackage{mathptmx}
\usepackage{etoolbox}
\usepackage[normalem]{ulem}
\usepackage{xr}
\usepackage{amsmath}
\usepackage{amssymb}
\usepackage{amsfonts}
\usepackage{array}
\usepackage{multirow}

\newcommand{\be}{\begin{equation}}
\newcommand{\ee}{\end{equation}}
\newcommand{\baln}{\begin{aligned}}
\newcommand{\ealn}{\end{aligned}}
\newcommand{\bal}{\begin{align}}
\newcommand{\eal}{\end{align}}
\newcommand{\bra}[1]{{\langle #1 |}}
\newcommand{\ket}[1]{{| #1 \rangle}}

\newcommand{\MoI}{MoI$_3$\xspace}
\newcommand{\MoBr}{MoBr$_3$\xspace}
\newcommand{\MoX}{MoX$_3$\xspace}

\newcommand{\av}{{\bf a}}

\newcommand{\cv}{{\bf c}}

\newcommand{\kv}{{\bf k}}

\newcommand{\Sv}{{\bf S}}
\newcommand{\Stot}{S_{\rm tot}}

\newcommand{\Rv}{{\bf R}}

\newcommand{\Jc}{{\mathcal{J}}}

\begin{document}

\title{Quantum Spin Singlet Ground State with Large Excitation Gap in 
the $S=3/2$ quasi-1D van der Waals materials MoX$_3$ (X = I, Br)}
\author{Jordan Teeter}
\affiliation{ 
Department of Materials Science and Engineering, University of California, Los Angeles, California 90095 USA}
\affiliation{California NanoSystems Institute, University of California, Los Angeles, California 90095 USA}
\author{Topojit Debnath}%
\affiliation{ Department of Electrical and Computer Engineering,
University of California, Riverside, CA, 92521, USA}
\author{Harshil Goyal}%
\affiliation{Department of Electrical and Computer Engineering, Auburn University, Auburn, Alabama 36849 USA}
\author{Md. Sabbir Hossen Bijoy}
\affiliation{Materials Research and Education Center, Department of Mechanical Engineering, Auburn University, Auburn, Alabama 36849 USA}
\author{Maedeh Taheri}
\affiliation{ 
Department of Materials Science and Engineering, University of California, Los Angeles, California 90095 USA}
\affiliation{California NanoSystems Institute, University of California, Los Angeles, California 90095 USA}
\author{Nicholas R. Sesing}
\affiliation{Department of Chemistry, University of Georgia, Athens, Georgia 30602 USA}
\author{Clayton Mantz}
\affiliation{Department of Chemistry, University of Georgia, Athens, Georgia 30602 USA}
\author{Josemaria Garcia}
\affiliation{Department of Chemistry, University of Georgia, Athens, Georgia 30602 USA}
\author{Mathew B. Stone}
\affiliation{Neutron Scattering Division, Oak Ridge National Laboratory, Oak Ridge, Tennessee 37830, USA}
\author{Colin Sarkis}
\affiliation{Neutron Scattering Division, Oak Ridge National Laboratory, Oak Ridge, Tennessee 37830, USA}
\author{Do Hoon Kiem}
\affiliation{Materials Science and Technology Division, Oak Ridge National Laboratory, Oak Ridge, Tennessee 37830, USA}
\author{Sohee Kwon}
\affiliation{Materials Science and Technology Division, Oak Ridge National Laboratory, Oak Ridge, Tennessee 37830, USA}
\author{Mina Yoon}
\affiliation{Materials Science and Technology Division, Oak Ridge National Laboratory, Oak Ridge, Tennessee 37830, USA}
\author{Fariborz Kargar}
\affiliation{Materials Research and Education Center, Department of Mechanical Engineering, Auburn University, Auburn, Alabama 36849 USA}
\author{Kirill Shtengel}
\affiliation{Department of Physics and Astronomy,
University of California, Riverside, CA, 92521, USA}
\author{Tina T. Salguero}
\affiliation{Department of Chemistry, University of Georgia, Athens, Georgia 30602 USA}
\author{Alexander A. Balandin}%
\email[Corresponding author: ]{balandin@seas.ucla.edu}
\affiliation{ 
Department of Materials Science and Engineering, University of California, Los Angeles, California 90095 USA}
\affiliation{California NanoSystems Institute, University of California, Los Angeles, California 90095 USA}
\affiliation{Center for Quantum Science and Engineering, University of California, Los Angeles, California 90095 USA}
\author{Roger K. Lake}
\email[Corresponding author: ]{rlake@ucr.edu}
\affiliation{ 
Department of Electrical and Computer Engineering,
University of California, Riverside, CA, 92521, USA}
%


\date{\today}

\begin{abstract}
\centerline{\textbf{Abstract}}
\vspace{0.5em}
We report the first experimental evidence for a quantum-disordered spin-singlet ground 
state and gapped triplon excitations in quasi-one-dimensional 
van der Waals materials.
The insulating van der Waals trihalides \MoI and \MoBr
are analyzed in a combined
thermodynamic, spectroscopic, and theoretical investigation. 
These materials contain structurally dimerized atomic chains of face-sharing octahedra 
with alternating short and long Mo--Mo separations along the chain direction.
Temperature-dependent bulk magnetic susceptibility measurements reveal a broad 
thermally activated response consistent with a large spin gap.
Inelastic neutron scattering (INS) spectra 
provide direct evidence of a collective spin gap at an exceptionally large energy transfer of $\Delta \approx 30\text{ meV}$ ($\approx 348\text{ K}$) for \MoBr. 
The absence of low-energy magnetic spectral weight and the activated susceptibility are 
inconsistent with a conventional classical antiferromagnetic ground state. 
Instead, they point to 
a spin-singlet ground state with a large singlet-triplet excitation gap.
To identify the microscopic origin of this gap, 
we perform density matrix renormalization group (DMRG) 
calculations on one-dimensional spin chains using exchange constants derived from 
density functional theory (DFT).
The dominance of the single dimer exchange term allows the inter-dimer coupling 
along the chain and between chains to be treated using perturbation theory.
Intrachain exchange lowers the isolated-dimer triplet energy by producing a dispersive triplon band. 
INS data show the triplon bandwidth to be
$\sim 20$ meV.
Interchain coupling produces negligible additional bandwidth on the order of 1 meV.
The INS measurements directly establish quantum spin-singlet and triplon 
physics in \MoBr with a large singlet-triplet gap on the scale of room temperature, 
while the susceptibility and theoretical results indicate closely analogous 
behavior in \MoI.
\end{abstract}

\maketitle

\section{\label{sec:Intro}Introduction}
Reduced dimensionality provides a powerful route for stabilizing quantum states that are 
difficult to realize in conventional three-dimensional solids. 
In one-dimensional (1D) systems, 
enhanced quantum fluctuations strongly suppress classical long-range order and make the ground 
state highly sensitive to spin magnitude, exchange topology, anisotropy, and structural 
dimerization. 
Quasi-one-dimensional (Q1D) van der Waals (vdW) materials provide a particularly 
useful setting for exploring this physics because strongly bonded atomic chains are separated by 
much weaker interchain interactions, 
allowing the intrinsic chain physics to remain visible in 
bulk crystals and, potentially, in exfoliated few-chain or single-chain
limits
\cite{2022_1D_Review_Mat_Today,
2018_Single_Chain_TMT_Sci,
2021_NbTe3_VTe3_TiTe3_in_CNT_JACS,
2025_Tunable_Synthesis_VxTey_Magnets_CNTs_NCom,
2025_High-Spin_State_CrX2_JACS,
2025_Qu_Sp_Dyn_1D_vdW_JACS,
2024_1D_vdW_AAB_AdvMat, Single-Chain_encap}.

Magnetic Q1D-vdW materials are therefore natural platforms for testing basic ideas of 
one-dimensional quantum spin physics. 
The theoretical foundation begins with the spin-(1/2) 
Heisenberg antiferromagnetic chain, 
where Bethe's exact solution revealed a quantum-disordered 
ground state and gapless excitations \cite{bethe1931}.
Later work established that half-integer spin chains  are generically 
gapless in the uniform limit \cite{lieb1961}, 
while Haldane predicted that integer-spin antiferromagnetic chains possess a 
finite excitation gap \cite{haldane1983}.
These canonical results are modified by perturbations that are common in real materials. 
In particular, bond alternation can open a singlet-triplet gap in half-integer 
spin chains and can drive integer-spin chains from the Haldane regime into a 
topologically trivial dimer-singlet phase
\cite{1994_S1_Dimerized_JPSJ,
1996_S1_Dimerization_Anis_Phase_Diag_JPSJ,
1996_S_3-2_Dimerized,
1997_Phase_Trans_Dimerized_spin_chains_Yamamoto_PRB,
1999_S-half_dimerized_Tennant_PRB}.
Interchain exchange then determines whether the gapped 
1D state survives in the bulk or gives way to a more classical ordered state.

The molybdenum trihalides \MoX (X = I, Br) 
occupy a distinctive position within this landscape. 
Their crystal structures contain chains of face-sharing
MoX$_6$ octahedra with alternating short and long Mo--Mo separations along the chain direction \cite{1972_MoBr3_Structure,2004_RuBr3_MoBr3_ChemMat}. 
The formal Mo(III) valence corresponds to a d$^3$ configuration and a local spin
$S=3/2$.
This combination of high spin, strong structural dimerization, and weak interchain coupling 
suggests that \MoX
may realize a strongly dimerized high-spin quantum spin chain.
However, prior theoretical studies of \MoX
have largely treated these materials within classical magnetic ground-state 
pictures
\cite{2024_Topo_Mag_AFM_Chain_PRB,
2024_MoI3_Spin_Chains_JMMM,
2025_Toroidic_PRB,
2025_Strain_Tunable_PRM}.
Such descriptions naturally lead to ordered spin configurations and spin-wave-like excitations, but they do not capture the possibility that the true ground state is a nonmagnetic 
singlet-dimer phase adiabatically connected to the isolated-dimer
limit.
A further ambiguity arises because DFT calculations yield local Mo moments slightly above 
$2\mu_B$, smaller than the $3\mu_B$ spin-only value expected for an ionic $S=3/2$ d$^3$ ion. 
Below we treat this as an effective-spin question and show that 
an $S=1$ model with DFT-derived exchange parameters 
overestimates the singlet-triplet gaps by a factor of two relative 
to experiment.

Existing experimental information has also left the magnetic ground state unresolved. 
X-ray photoelectron spectroscopy suggests predominantly Mo(III) character in \MoI, and prior Raman measurements revealed low-temperature features that were interpreted in terms 
of magnetic excitations
\cite{2020_MoI3_photodetectors, 2026_MoI3_pellet, 2022_MoI3_Fari_APL}.
These observations indicate that spin degrees of freedom are active, 
but they do not  distinguish between a classical antiferromagnetic state and 
a gapped singlet ground  state. 
Resolving this distinction requires thermodynamic and 
spectroscopic probes of the low-energy magnetic spectrum, together with a theoretical treatment that includes the quantum nature 
of the spin chain.

In this work, we determine the 
spin ground state and excitation gap of 
\MoI and \MoBr through a combined experimental and theoretical study. 
Magnetic susceptibility measurements reveal a thermally activated response consistent 
with a large spin gap, and inelastic neutron scattering on
\MoBr directly resolves a magnetic excitation gap
$\Delta \approx 30$ meV.
These observations are inconsistent with a conventional classical antiferromagnetic 
ground state  with low-energy spin-wave excitations. 
Instead, they point to a nonmagnetic spin-singlet ground 
state and a large singlet-triplet excitation gap. 
To identify the microscopic origin of this gap, 
we combine density-functional-theory-derived exchange constants with density matrix 
renormalization group calculations for single $S=3/2$ chains 
and first-order triplon perturbation 
theory for the full three-dimensional exchange network. 
The resulting picture is that the short 
Mo--Mo bond forms a strong antiferromagnetic dimer, 
while weaker intra- and interchain exchanges 
disperse the triplet excitation, but do not close the gap.
\MoX therefore realizes a strongly dimerized, high-spin, quasi-one-dimensional quantum spin 
system with a robust singlet ground state and a large triplon gap on the scale of room 
temperature.

\section{Methods}
\label{sec:methods}
%
This section provides
an overview of the experimental and theoretical
methods used in this work.
The detailed procedures and parameters required to reproduce
the experimental and theoretical results are provided in 
Appendices \ref{app:Exp_methods} and~\ref{app:Comp_methods},
respectively.
Additional data and extended discussion are provided in the
Supplemental Material (SM)~\cite{SM}, together with supporting
references~\cite{2013_Qu_Class_Xover_PRB,
Gruber1983,Schlenker_1986,MZoller_2020,Sandeep_2015,
Sharmistha_2024,1981_Mag_solid_O2,2017_H_T_Phase_diag_O2,
Nomura_2017,7405275,QD_MPMS_Notes}.

Single crystals of \MoI and \MoBr were grown by chemical vapor transport (CVT)
and characterized by x-ray diffraction (XRD), scanning electron microscopy (SEM),
and energy-dispersive x-ray spectroscopy (EDS).
X-ray photoelectron spectroscopy (XPS) was performed on freshly exfoliated
crystals to minimize surface oxidation.
Heat capacity was measured between 1.8 and 100 K using the thermal
relaxation method in a Quantum Design DynaCool Physical Property Measurement System (PPMS).
Magnetic susceptibility was measured between 1.8 and 300 K in an
applied field of 0.5 T using a 
Quantum Design Magnetic Property Measurement System 3 (MPMS3) magnetometer.
The sample-holder contribution was measured separately and subtracted.
INS measurements were performed on
approximately 1 g of \MoBr powder using the SEQUOIA chopper
spectrometer at the Spallation Neutron Source \cite{SEQUOIA}.
The incident neutron energy was $E_i=80$ meV, and spectra were
collected at 5, 60, 150, and 300 K.
Empty-can measurements were used for background subtraction, and
vanadium standards were used to calibrate the detector efficiencies.
Experimental procedures, sample preparation, and measurement
parameters are given in Appendix~\ref{app:Exp_methods}.
%
%

DFT calculations were performed using the
Vienna Ab initio Simulation Package 
(VASP) \cite{kresse1993ab,kresse1996efficient} with the 
Perdew, Burke, and Ernzerhof (PBE)
exchange-correlation functional  \cite{perdew1996generalized},
a van der Waals correction \cite{grimme2010consistent}, 
and on-site Coulomb interactions treated within DFT+$U$.
Magnetic exchange interactions were obtained using the magnetic-force
linear-response method implemented in Jx \cite{yoon2020jx}, with
Wannier orbitals generated using Wannier90
\cite{pizzi2020wannier90}.
Magnetocrystalline anisotropy was determined from spin-orbit-coupled
DFT calculations.
The resulting interactions were used to construct an $S=3/2$
Heisenberg model.
The singlet-triplet gaps of finite, open spin chains were calculated
using DMRG as implemented
in TeNPy \cite{TeNPy}, retaining the four intrachain exchange
interactions, both with and without single-ion anisotropy.
The thermodynamic-limit gaps were obtained by finite-size
extrapolation.
The triplon dispersion was evaluated by first-order perturbation
theory about the isolated-dimer limit.
Phonon scattering for comparison with the neutron data was calculated
from DFT force constants using INSPIRED \cite{HAN2024109288} and
combined with a powder-averaged magnetic scattering model.
The exchange parameters and the technical details of the numerical procedures 
and convergence criteria are given in Appendix~\ref{app:Comp_methods}.

\section{Experimental Results}
\label{sec:exp}
\subsection{Materials Characterization, XPS, and Heat Capacity}
\label{sec:Matcharc}
\begin{figure}
\centering
\includegraphics[width=3.5in]{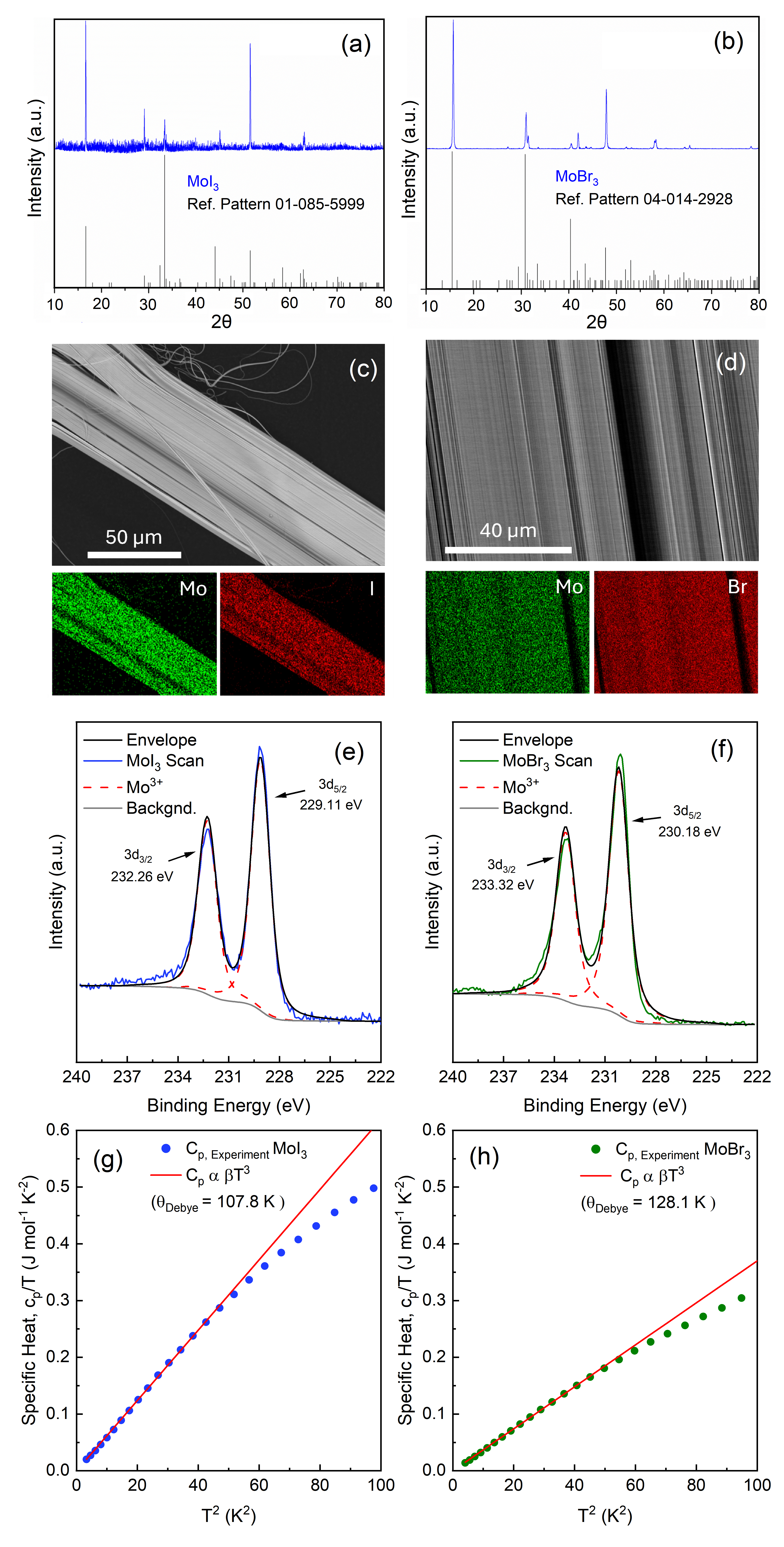}
\caption{
(a, b) X-ray diffraction patterns for (a) \MoI and (b) \MoBr. 
(c, d) Scanning electron microscopy images of exfoliated samples with corresponding 
energy dispersive spectroscopy maps for (c) \MoI and (d) \MoBr. (e, f) High-resolution x-ray photoelectron spectra of the Mo 3d region for mechanically exfoliated (e) \MoI and
(f) \MoBr. The experimental data are shown with fits to the Mo 3d spin-orbit doublet. (g, h) $C_p/T$ vs $T^2$ from 1.8 K to 9.8 K for  
(g) \MoI and (h) \MoBr . Blue and green points represent experimental data, while red lines represent the Debye approximation at low temperatures, where $C_p = \beta T^3$, with $\beta = 0.0062$ and $0.0037$ for \MoI and \MoBr respectively. 
}
\label{fig:struct}
\end{figure}

%
The crystalline phase and quality of the \MoI and \MoBr crystals were verified by XRD. 
As shown in Figs. \ref{fig:struct}(a,b), the recorded diffraction patterns match their corresponding International Centre for Diffraction Data (ICDD) reference patterns, 
confirming the \textit{Pmmn} space group for \MoI and \textit{Pnmm} space group for \MoBr, in agreement with prior reports
\cite{2022_MoI3_Fari_APL,1972_MoBr3_Structure,2004_RuBr3_MoBr3_ChemMat}. 
For ease of comparison in this work, 
the unit cell of \MoBr has been 
transformed from the \textit{Pnmm} space group to the 
\textit{Pmmn} space group, 
so the chain direction is along the $b$-axis for both \MoI and \MoBr. 
The experimental lattice constants
of \MoI are $a=12.32$ \AA, $b=6.41$ \AA,
and $c=7.12$ {\AA}  \cite{2022_MoI3_Fari_APL}. 
For \MoBr, after transforming to \textit{Pmmn}, they are
$a=11.44$  \AA, $b=6.08$ \AA, and $c=6.61$ {\AA} 
\cite{2004_RuBr3_MoBr3_ChemMat}.
Due to the differing ionic radii of halogen species ($X$ = I, Br), the degree of dimerization varies between the two crystals. 
\MoI has alternating Mo-Mo distances of 2.88 Å and 3.52 Å, and \MoBr has distances of 2.87 Å and 3.20 Å.
SEM imaging (Figs. \ref{fig:struct}(c,d)) reveals flexible, fibrous crystals having 
a broad diameter distribution ranging from $\sim 0.5$ mm to $\sim 50$ nm, demonstrating their highly facile cleavage.
Additional SEM images are shown in Fig. S1 of the SM \cite{SM}.
EDS yields experimental atomic ratios of Mo\,:\,I = 24.7\,:\,75.3 for \MoI 
and Mo\,:\,Br = 25.0\,:\,75.0 for \MoBr, 
consistent with the expected 1\,:\,3 stoichiometric ratios. Furthermore, elemental
EDS mapping 
confirms a highly uniform spatial distribution of constituents across the crystals.

High-resolution X-ray photoelectron spectra of the Mo 3d region were collected from 
mechanically exfoliated samples to assess the chemical state of molybdenum in \MoI (blue) 
and \MoBr (green) (Figs. \ref{fig:struct}(e,f)). The spectra were analyzed by fitting 
constrained Mo 3d spin–orbit doublets, yielding Mo 3d$_{5/2}$ binding energies of 229.11 
eV for \MoI and 230.18 eV for \MoBr. Notably, neither spectrum exhibits higher binding 
energy shoulders or additional components attributable to Mo(IV), Mo(V), or Mo(VI) 
species, indicating that little, if any, oxidized molybdenum is present at the surfaces of 
these exfoliated samples. 
This observation contrasts with published XPS data for \MoI containing distinct higher binding energy features \cite{2026_MoI3_pellet, 2020_MoI3_photodetectors}.

The measured Mo 3d binding energies for \MoI and \MoBr fall within the range reported for 
low valent molybdenum halides. Reported Mo 3d$_{5/2}$ binding energies for \MoI of 228.4 
and $\sim$228.9 eV \cite{2026_MoI3_pellet, 2020_MoI3_photodetectors} are comparable to the 
values measured here. Further comparison can be made with Mo(II) cluster halides. These 
compounds exhibit comparable Mo 3d$_{5/2}$ binding energies of 229.8 eV for 
Mo$_6$Br$_{12}$ \cite{2023_Mo6Br12}, 228.7 eV for Mo$_6$Cl$_{12}$ \cite{1974_Mo6Cl12}, and 
228.4 eV for Mo$_6$I$_{12}$ \cite{1978_Mo6I12}. Relative to the corresponding Mo(II) 
cluster halides, the Mo 3d$_{5/2}$ binding energies for \MoI and \MoBr increase by 
approximately 0.7 eV and 0.4 eV, respectively. Although ligand environments and metal-
metal interactions also influence binding energies, these shifts are consistent with the 
higher formal oxidation state of Mo(III) in \MoI and \MoBr.

Survey spectra confirming the elemental composition of each sample, as well as the 
corresponding high-resolution halide core-level spectra, are provided in the SM (Figs. S8 
and S9). Both halide spectra are well fit by a single spin–orbit doublet, as expected for 
the 1D chain structures of \MoI and \MoBr having all halides in a single chemical 
environment (bridging Mo centers). Consistent with this interpretation, the Br 3d$_{5/2}$ 
binding energy of 69.53 eV for \MoBr is in line with that for the “inner” bridging 
bromides in Mo$_6$Br$_{12}$ at 70.7 eV \cite{2023_Mo6Br12}.
Collectively, these XPS results provide spectroscopic evidence that the molybdenum in \MoI 
and \MoBr is present in the formally expected Mo(III) chemical state and that each 
compound exhibits a single halide chemical environment, in agreement with the structures 
determined by single crystal x-ray diffraction 
\cite{1972_MoBr3_Structure,2022_MoI3_Fari_APL,2004_RuBr3_MoBr3_ChemMat}.

To probe the differences in the weak interchain van der Waals bonding between \MoX crystals, 
heat capacity ($C_p$) measurements were also performed at low temperatures for randomly 
oriented  \MoI and \MoBr  samples (Figs. \ref{fig:struct}(g, h)). Given the insulating 
nature of these materials, electronic contributions to the heat capacity are expected to be 
negligible. At  low temperatures, bulk \MoI (blue) and \MoBr (green) exhibit conventional cubic 
temperature dependence, $C_p/T = \beta T^2$, where $\beta = 12\pi^4 NR/5\theta_{D}^3$ (with $N = 
4$ representing the number of atoms per formula unit, $R$ the universal gas constant, and $
\theta_D$ being the Debye temperature). Fitting the data yields exceptionally small Debye 
temperatures of $\theta_D = 107.8\text{ K}$ for \MoI and $\theta_D = 128.1\text{ K}$ for \MoBr, which reflect an inherently soft lattice. In these Q1D crystals, the 
preservation of a $T^3$ power law at low temperatures indicates that the phonon spectrum remains 
effectively three-dimensional, likely dominated by soft, transverse interchain vibrational modes 
arising from weak van der Waals interactions \cite{1986_ZrSe3andZrS3, 2022_XiChen_ZrSe3}. The 
lower Debye temperature of \MoI relative to \MoBr could stem from the larger ionic radius and mass of 
iodine compared to bromine, which increases interchain separation. 
This structural variation possibly accounts for the difference between the wire-like morphology of \MoI compared to the more rigid and larger \MoBr crystals observed via SEM.

%
\subsection{Magnetic Susceptibility}
\label{sec:MagSus}
\begin{figure}
\centering
\includegraphics[width=3.5in]{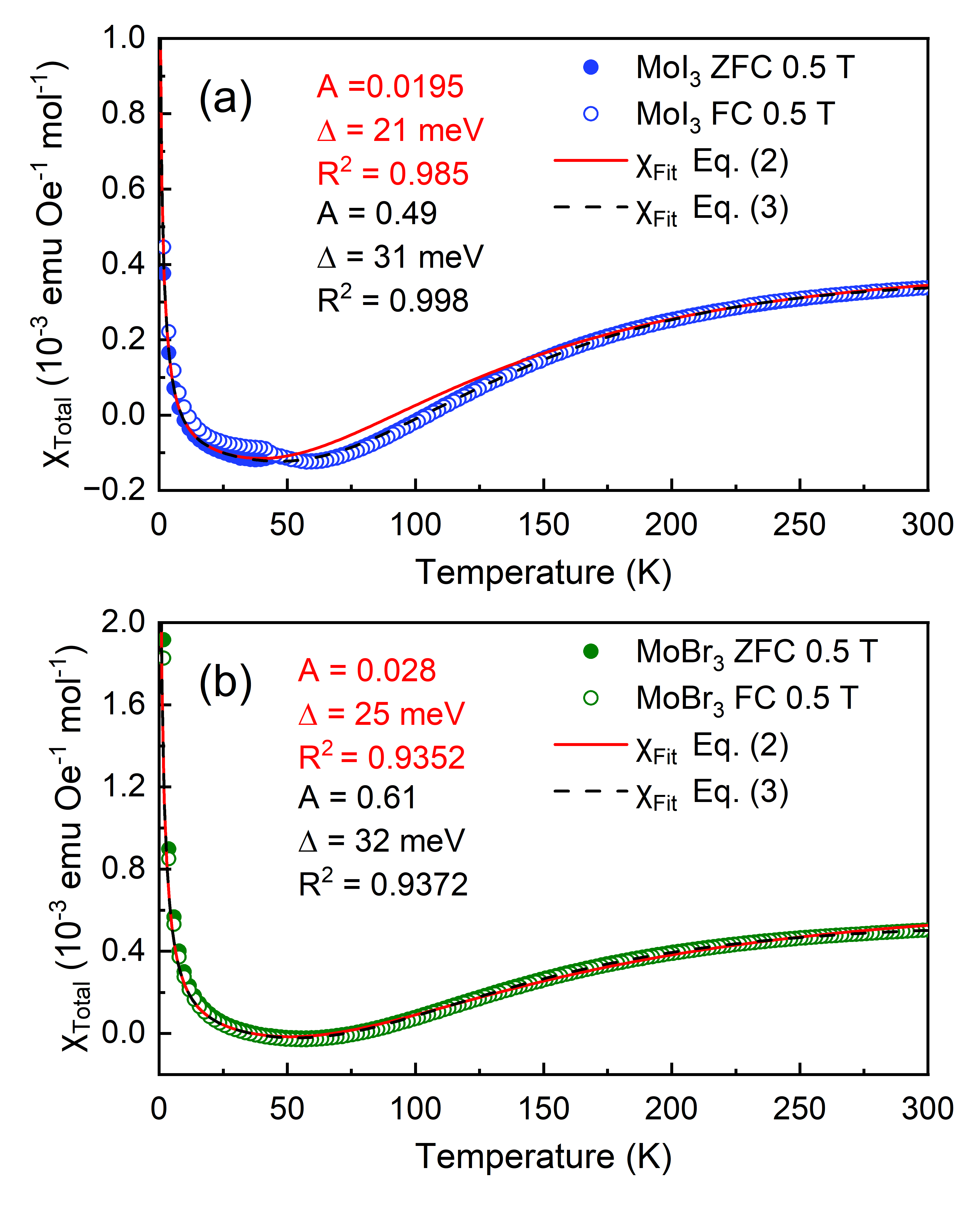}
\caption{
Magnetic susceptibility of \MoI and \MoBr.  
Zero field cooled (ZFC) and field cooled (FC) magnetic susceptibilities of 
(a) \MoI and (b) \MoBr from 1.8 K to 300 K at 0.5 T fit to Eq. (\ref{eq:chi_emp0}) using forms (\ref{eq:chi_t1_Haldane}) and (\ref{eq:chi_t2})
for $\Theta(T)$
}
\label{fig:chi.png}
\end{figure}

%
Turning to the magnetic properties, magnetic susceptibility was measured as a function of temperature for 
randomly oriented \MoI and \MoBr crystals, 
with the results shown in Figs. \ref{fig:chi.png}(a, b). 
The two most notable features are the low temperature Curie tail
resulting from free paramagnetic spins, defects, or vacancies
\cite{1979_Exchange-coupled_pair_PRB}
and the high-temperature thermally activated broad peak
associated with a spin gap $\Delta$
\cite{2000_Thermo_S1-2_AFM_Chains_PRB,
2011_Spin_Dyn_Organic_Lin_Chains_PRB}.

To extract the spin gap from the susceptibility, we fit the
experimental data to a generic form of the susceptibility for a gapped 
1D spectrum,
\be
\chi(T) = A \Theta(T) + \frac{C_{\rm tail}}{T} + \chi_0 .
\label{eq:chi_emp0}
\ee
In Eq. (\ref{eq:chi_emp0}), the first term represents
the contribution from the thermal excitation across the spin gap, with $A$ 
representing an amplitude and $\Theta(T)$ 
the activated temperature factor.
We will consider two different forms for $\Theta(T)$ below.
The second term accounts for the low-temperature Curie tail resulting from
free spins.
The third term, $\chi_0$, represents
the temperature-independent 
ionic core diamagnetism, a positive Van Vleck term, 
and any background contributions.
The sample-holder background is removed by the 
empty-paddle subtraction described in
Appendix \ref{app:Exp_methods}.
The $C_{\rm tail}$ and $\chi_0$ terms have been quantitatively extracted from the inverse 
susceptibility at low temperatures by fitting to a Curie-Weiss law (see Fig. S4 of the SM).
The values are given in Table \ref{tab:chi_params}.

The fitted $\chi_0$ values are negative and of the expected size.
Summing tabulated ionic Pascal's constants~\cite{Bain2008} gives
$-1.75 \times 10^{-4}$ and $-1.27 \times 10^{-4}$~emu~Oe$^{-1}$mol$^{-1}$ for 
MoI$_3$ and MoBr$_3$, respectively.
The fitted values lie above these estimates in
both cases, the direction expected once the positive Van Vleck term
and the reduction of the ionic diamagnetism by covalency are included.

The values for $C_{\rm tail}$ provide an estimate for the 
percent of free spins in the samples.
The susceptibility of uncoupled spins is
$C/T$ where $C = N_A g^2 \mu_B^2 S(S+1)/(3 k_B)$.
With $g=2$ and $S=3/2$, $C=1.88$ emu Oe$^{-1}$ mol$^{-1}$ K.
The ratios
$C_{\rm tail}/C$ are $8.3 \times 10^{-4}$ for \MoI
and $2.0 \times 10^{-3}$ for \MoBr.
This indicates that 0.08\% (0.2\%) of the Mo sites fail to pair
into dimers for \MoI (\MoBr).
Note that $g$ is assumed rather than measured, and if the 
defect spins are chain end spins carrying $S=1/2$ or $S=1$, 
the percentages increase by $5$ or $15/8$, respectively.

The first form of $\Theta(T)$ that we consider 
is one originally derived for a 1D gapped phase \cite{1995_Chi_Fit_JMMM, Random_exchange_Haldane},
\be
\Theta_1(T) = \frac{1}{\sqrt{T}} e^{-\Delta/k_BT}  .
\label{eq:chi_t1_Haldane}
\ee
The $1/\sqrt{T}$ prefactor arises from the 1D density of states of the excited
triplon band.
Under large dimerization, an $S=3/2$ chain approaches the (3,0)
valence bond state 
(three valence bonds within each dimer and none between dimers),
equivalent to independent dimers
\cite{1997_Phase_Trans_Dimerized_spin_chains_Yamamoto_PRB}.
The density of states transitions from 1D to zero-dimensional (0D), 
and the prefactor transitions from
$1/\sqrt{T}$ to $1/T$.
Therefore, the second form of $\Theta(T)$ considered is
\be
\Theta_2(T) = \frac{1}{T} e^{-\Delta/k_BT}  ,
\label{eq:chi_t2}
\ee
which is the leading term in a low-temperature 
approximation to the exact expression for $S=3/2$ dimer given 
by Eq. (\ref{eq:chi_B_dimer}).
%

%

The resulting best-fit curves to the susceptibility data are shown by the solid red and black dashed curves
in Fig. \ref{fig:chi.png}(a,b).
Eq. (\ref{eq:chi_emp0}) provides a good fit to both the low temperature regions of the
susceptibility curves dominated by the Curie tail from the free spins and the high temperature regions with Eqs. (\ref{eq:chi_t1_Haldane}) and (\ref{eq:chi_t2}) fitting the thermally
activated spin gap, respectively.
Eq. (\ref{eq:chi_t1_Haldane}) yields spin gaps of $\Delta = 21$ meV for \MoI and $\Delta = 25$ meV for \MoBr, while Eq. (\ref{eq:chi_t2}) yields spin gaps of $\Delta = 31$ meV for \MoI and $\Delta = 32$ meV for \MoBr.
 Fitting the experimental data in this way provides estimates of 
 the spin gaps for a system with a wide, dispersive triplon band
 with a 1D density of states and for a strongly dimerized
 chain with a narrow triplon band approaching a 0D density of states.
 These different regimes are analyzed in Appendix \ref{sec:T-prefactor}.
 The fitting parameters are provided in Table \ref{tab:chi_params}.

There is one other small feature in the \MoI susceptibility curve (Fig. \ref{fig:chi.png}(a))
consisting of a small cusp with splitting between the 
ZFC and FC curves observed at  $T \approx 44$ K. 
A much smaller feature at $T \approx 41$ K is also observed in the \MoBr data 
(Fig. \ref{fig:chi.png}(b)). 
We attribute these features to an extrinsic
oxygen-related origin with further discussion
provided in the SM.
\begin{table}
\centering
\caption{
Parameters for the fitted curves in Fig. \ref{fig:chi.png}.
}
\begin{tabular}{|l|c|c|}
\hline
Parameter & \MoI & \MoBr \\
\hline
\hline
$\Delta$$_{Eq. 2}$ (meV) & $21$ & $25$\\
\hline
$\Delta$$_{Eq. 3}$ (meV) & $31$ & $32$\\
\hline
$A_{Eq.2}$ (emu Oe$^{-1}$ mol$^{-1}$ K$^{1/2}$) & $1.95\times 10^{-2}$ & $2.8\times 10^{-2}$\\
\hline
$A_{Eq.3}$ (emu Oe$^{-1}$ mol$^{-1}$ K) & $0.49$ & $0.61$\\
\hline
$C_{\rm tail}$  (emu Oe$^{-1}$ mol$^{-1}$ K) & $1.56 \times 10^{-3}$ & $3.7 \times 10^{-3}$\\
\hline
$\chi_0$ (emu Oe$^{-1}$ mol$^{-1}$) & $-1.6 \times 10^{-4}$ & $-1.0 \times 10^{-4}$\\
\hline
\end{tabular}
\label{tab:chi_params}
\end{table}
%
\subsection{Inelastic Neutron Scattering}
\label{sec:INSexp}
%
To determine the extent of the magnetic excitation spectrum and to directly 
probe the spin gap $\Delta$ inferred from the magnetic  susceptibility, 
INS was performed on \MoBr.
\MoBr was selected because its high-yield synthesis provides the sample mass
required for INS.
%
%
Approximately 1 g of purified powder 
(see SM for the purification procedure)
was measured on the SEQUOIA chopper spectrometer at the Spallation Neutron 
Source \cite{SEQUOIA}
with an incident neutron energy $E_i = 80$ meV
at $T$ = 5, 60, 150, and 300 K.
The data shown have an empty-can background subtracted. 
Further experimental detail is given in App. \ref{app:Exp_methods}.

\begin{figure}
\centering
\includegraphics[width=3.5in]{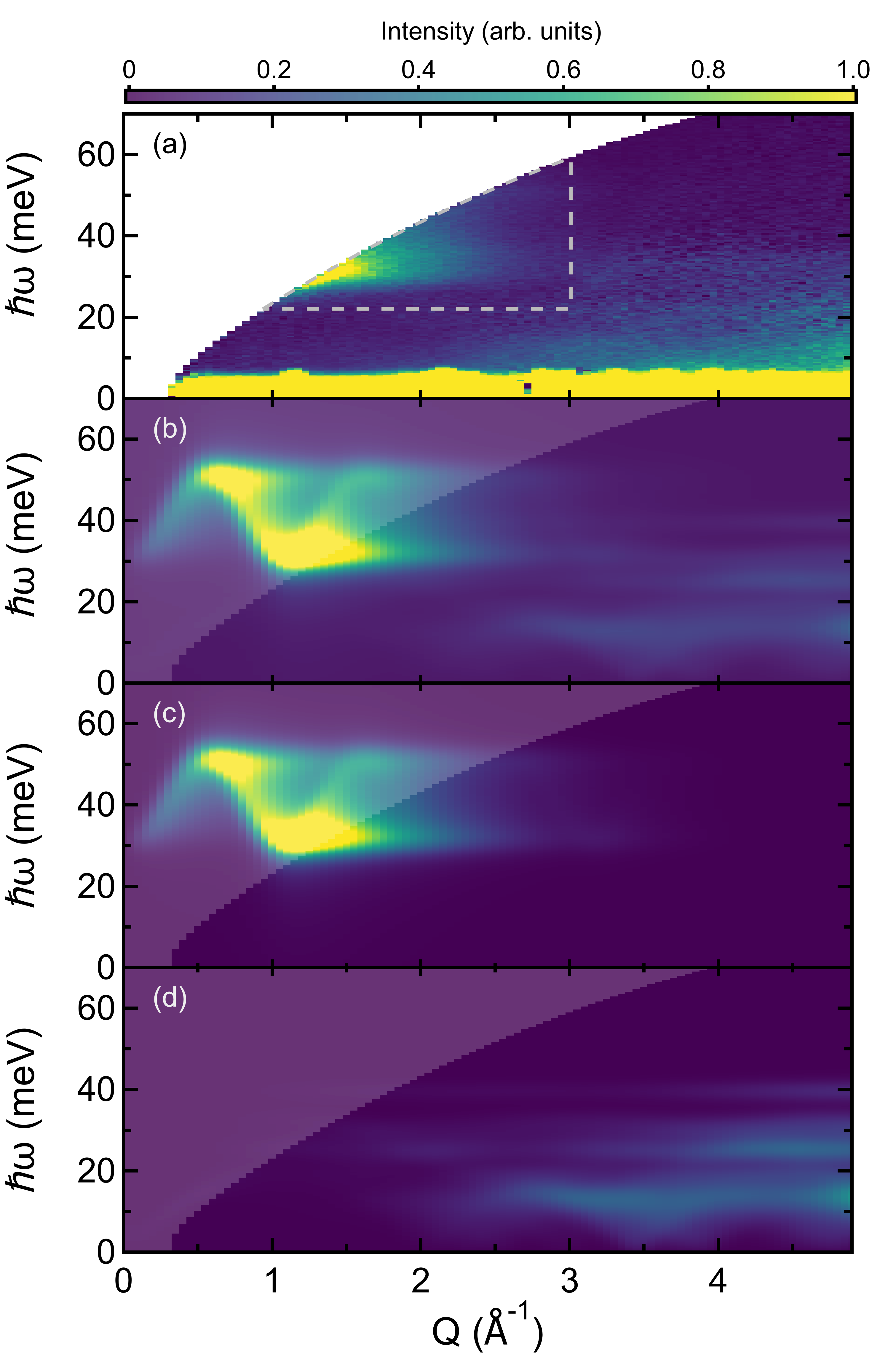}
\caption{
(a) Background-subtracted false color plot of the experimentally measured inelastic neutron scattering intensity for \MoBr at $T = 5$ K as described in the text. (b) Model of scattering intensity including both alternating chain spin and phonon contributions to the spectrum.  (c) Alternating chain spin portion of the modeled scattering intensity.  (d) Phonon portion of the modeled scattering intensity.   The model calculations are plotted with a different transparency value for data outside of the kinematically allowed coordinates for the measurement.  Phonon calculation in panel (d) is shown with twice its calculated relative scattering intensity.  Dashed line in (a) corresponds to the range of the measured data used to refine the alternating chain exchange constants as described in the text.
}
\label{fig:neutron}
\end{figure}

\begin{figure}
\centering
\includegraphics[width=3.5in]{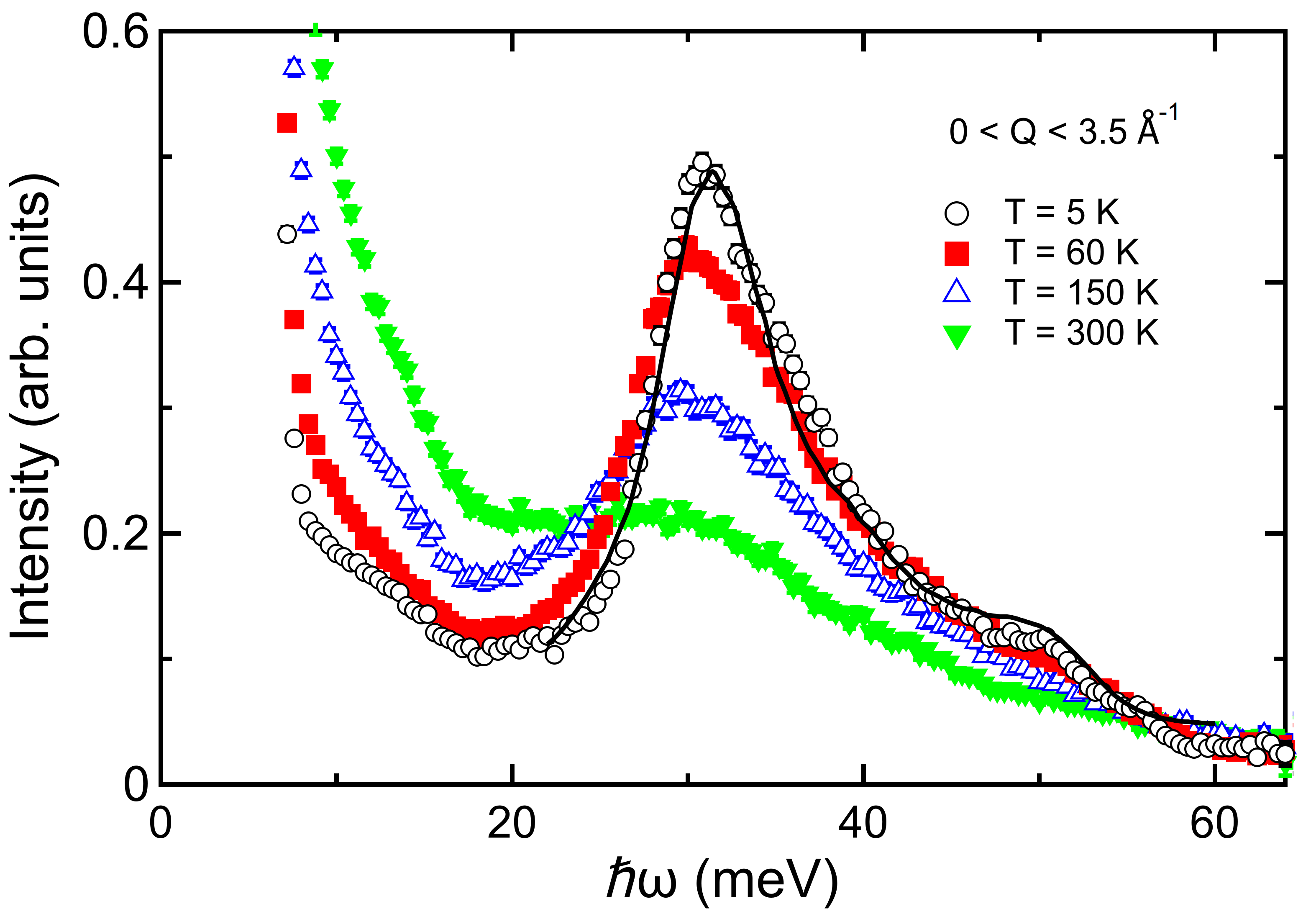}
\caption{
Background-subtracted integrated $Q$ scans for  $Q \in [0,3.5]$ Å$^{-1}$ at temperature $T = 5$~K, $60$~K, $150$~K and $300$~K.  
The solid line is the fit to the alternating chain model described in the text plotted over the range of energy transfers that were used for the fit.
}
\label{fig:neutronIvsE}
\end{figure}

Fig. \ref{fig:neutron}(a) illustrates the background subtracted measured scattering intensity as a function of wave-vector, $Q$, and energy transfer, $\hbar\omega$, for $T=5$~K.  Data have been normalized to be proportional to the dynamic scattering function $S(Q,\omega)$.  At this base temperature, \MoBr occupies its nonmagnetic singlet ground state, allowing incident neutrons to populate the  lowest-lying magnetic triplet state. 
Magnetic excitations are clearly resolved for lower values of wave-vector transfer for energy transfers between approximately 30 and 55 meV.   A well defined gap energy of $\Delta \approx$ 30 meV ($\approx$ 348 K) is seen in the data at $Q\approx 1.5$~\AA$^{-1}$. The scattering intensity reaches its maximum in the low-$Q$ regime and decreases towards the higher $Q$ regime. This intensity trend follows the characteristic  magnetic form factor expected for localized magnetic excitations. The large scattering intensity near zero energy transfer is due to the elastically scattered neutrons.
Conversely, phonon excitations emerge with increasing intensity in the high-$Q$ regime, behaving in accordance with the phonon structure factor. 
To evaluate the thermal evolution of this spin gap, Fig. \ref{fig:neutronIvsE} displays the integrated-$Q$ cuts over the range $Q \in [0,3.5]$ Å$^{-1}$  for the background-subtracted spectra at $T$ = 5 K, 60 K, 150 K, and 300 K. These data indicate that the magnetic excitations extend between approximately 30 and 55 meV energy transfer. As the temperature increases, the peak intensity in the vicinity of the spin gap decreases and broadens. This behavior is consistent with the expected thermodynamic temperature dependence, as the overall transition probability falls due to the thermal depopulation of the singlet ground state and the corresponding population of the excited states \cite{gapTdep}.

Comparing the directly measured gap energy of $\Delta$ = 30 meV to the $\Delta$ values extracted from the magnetic susceptibility models for \MoBr 
suggests that \MoBr lies closer to the isolated-dimer
limit.
In the subsequent section, we 
theoretically analyze this system.

%
%

\section{Theoretical Analysis}
\label{sec:Theory}
We now use DFT-derived exchange parameters, DMRG, 
and strong-dimer perturbation theory to determine whether the 
experimentally observed response is consistent with a robust 
singlet-dimer phase. 
Because the numerical exchange constants depend on the treatment of 
electronic correlations, 
our focus is on qualitative features that remain unchanged across the range of $U_{\mathrm{eff}}$ considered, i.e.,
the dominance of the intradimer antiferromagnetic exchange, 
the singlet character of the ground state, 
and the finite singlet–triplet gap.
Details of the DFT and DMRG calculations and the exchange parameters are given in Appendix~\ref{app:Comp_methods}.

\subsection{Spin Hamiltonian and isolated-dimer limit}
\label{sec:H}
%
We model the magnetic degrees of freedom by an isotropic Heisenberg Hamiltonian with exchange constants extracted from density-functional theory,
\begin{equation}
H =
\sum_{\langle i j\rangle}
J_{ij} \mathbf S_i\cdot \mathbf S_j ,
\label{eq:heisenberg_general}
\end{equation}
where $J_{ij}>0$ denotes antiferromagnetic exchange. 
Exchange constants extending out to 12 nearest neighbors are included in the analysis.
Their values are listed in 
Tables \ref{tab:MoBr3_params_vs_U} and \ref{tab:MoI3_params_vs_U} of 
Appendix \ref{app:Comp_methods} ordered by distance.
%
We use the convention that $\sum_{\langle ij\rangle}$ is a single sum over all bonds 
$\{ij\}$, so that the difference in energy between ferromagnetic and antiferromagnetic
alignment of two spins $\Sv_i$ and $\Sv_j$ is $2J_{ij} S^2$ where $S=3/2$.
This is consistent with the convention used in the DMRG code TeNPy \cite{TeNPy}.
For the analysis, the relevant 
classification is not the distance but the topology of the bond 
relative to the structural dimers.
We therefore introduce a separate set 
of topology-based exchange labels. 
The dominant intradimer exchange is 
denoted by $\mathcal{J}_1$, while the three intrachain, interdimer 
exchanges entering the one-triplon dispersion are denoted by 
$\mathcal{J}_{2}$, $\mathcal{J}_{3}$, and $\mathcal{J}_{4}$.
These correspond to the four nearest neighbor exchange couplings along a chain and are illustrated in 
Fig. \ref{fig:Js_MoBr3} of
Appendix \ref{app:Comp_methods}.
Interchain exchange terms are labeled as $J_m$.
The dominant exchange term $\Jc_1$ connects the two Mo ions in a structural dimer. 
Each Mo ion is treated as an $S=3/2$ spin, 
as expected for the formal d$^3$ Mo$^{3+}$ configuration. 
The remaining exchange terms that connect neighboring dimers along 
and between  chains are small compared with $\Jc_1$.
We will show that their effect on the singlet-triplet gap
can be included using perturbation theory.
Therefore, we begin by considering a single dimer.

The isolated dimer Hamiltonian is
\be
H_{\rm dimer}=\Jc_1 \mathbf S_1\cdot\mathbf S_2.
\label{eq:Hdimer}
\ee
For two $S=3/2$ spins, the total dimer spin can take the values
$S_{\rm d}=0,1,2,3$,
with energies
$E(S_{\rm d}) =
\frac{\Jc_1}{2}
\left[
S_{\rm d}(S_{\rm d}+1)-2S(S+1)
\right]$.
For $S=3/2$, this gives energies
$E(0)=-\frac{15}{4}\Jc_1$,
$E(1)=-\frac{11}{4}\Jc_1$,
$E(2)=-\frac{3}{4}\Jc_1$, and
$E(3)=\frac{9}{4}\Jc_1$
with degeneracies $1$, $3$, $5$, and $7$, respectively.
Thus, the isolated-dimer excitation energies relative to the singlet ground state are
$\Delta=\Jc_1$,
$\Delta_2=3\Jc_1$,
and
$\Delta_{3}=6\Jc_1$.
The lowest magnetic excitation is therefore a triplet separated from the singlet 
by the singlet-triplet excitation gap of $\Jc_1$. 
This establishes the zeroth-order origin of the large observed spin gap.

There is also a small, easy-plane (positive) single-ion anisotropy,
which for a single dimer reads
$H_D = D ( S_{1,z}^2 + S_{2,z}^2 )$.
Values for  $D$ lie in the range of 
$\Jc_1/100$ to $\Jc_1 / 45$, so that in all
cases, $D \ll \Jc_1$. 
The singlet and triplet matrix elements in the total spin 
basis $\ket{S_d, m}$ are 
$\bra{0,0} H_D \ket{0,0} = \frac{5}{2} D$, 
$\bra{1,\pm 1} H_D \ket{1,\pm 1} = \frac{17}{10} D$, 
and
$\bra{1,0} H_D \ket{1,0} = \frac{41}{10} D$. 
Therefore, to first order, the anisotropy reduces the excitation gap by
$\frac{4}{5} D$.

\subsection{Single-chain DMRG and the singlet-triplet gap}
\label{sec:DMRG}
%
To go beyond the isolated-dimer limit, we first considered a single chain 
with the four intrachain exchange values $\Jc_1$ - $\Jc_4$
obtained from DFT. 
The singlet-triplet gap was computed using DMRG calculations on finite open chains. 
%
%
The thermodynamic-limit gap was then obtained by extrapolating $\Delta_N$ 
as a polynomial in $1/N$.
The extrapolated DMRG gap is reduced from the isolated-dimer value $\Jc_1$
to approximately the 30 meV value determined by experiment.
This reduction is the result of the inter-dimer exchange ($\Jc_2$, $\Jc_3$, and $\Jc_{4}$).
It  does not destroy the singlet ground state, but it allows the localized triplet 
excitation to propagate along the chain, lowering the bottom of the triplet band.
The calculated DMRG ground state has 
$\langle \mathbf S_{\rm tot}^2\rangle=0$ while the first excited state has 
$\langle \mathbf S_{\rm tot}^2\rangle=2$, corresponding to $S_{\rm tot} = 1$, 
confirming that the computed gap is the singlet-triplet gap.
The ground-state entanglement entropy alternates between values close to 
$\ln 4 = \ln(2S+1)$ on cuts 
through the structural dimers and nearly zero on cuts between dimers, as expected for 
weakly coupled $S=3/2$ singlet dimers.
The change in the intradimer correlation $\langle \mathbf S_A\cdot\mathbf S_B\rangle$ 
between the first excited state and the ground state sums to approximately one (0.99), 
consistent with a single delocalized triplon excitation.
The spatial profile of the correlation difference is a finite-chain standing-wave version of a delocalized triplon.

\subsection{First-order triplon perturbation theory}
\label{sec:1st_order_triplon}

The strong hierarchy $\Jc_1 \gg |\Jc_n|, |J_m|$ motivates a perturbative treatment in which the 
product of dimer singlets is the unperturbed ground state and the lowest excited states are 
single triplons. 
A one-triplon basis state is written as a direct product over dimers,
\begin{equation}
\ket{n,\sigma,m}
=
\ket{t_{n,\sigma},m}
\prod_{(n',\sigma')\ne(n,\sigma)}
\ket{s_{n',\sigma'}}
\label{eq:one_triplon_basis}
\end{equation}
where $n$ labels the unit cell, $\sigma = {\rm I,\: II}$ 
labels the dimer within the unit cell, and 
$m\in\{-1,0,+1\}$ 
is the triplet magnetic quantum number.
$\ket{t_{n,\sigma},m}$
is a triplet state $\ket{S_d=1, m}$ located  in
unit cell $n$ on dimer $\sigma$.
Similarly, 
$\ket{s_{n,\sigma}}$
is a singlet state on dimer $n, \sigma$.

For a weak exchange bond connecting spin $\alpha\in\{A,B\}$ on dimer $\sigma'$, $n'$ to spin 
$\beta\in\{A,B\}$ on dimer $\sigma$, $n$, 
the coupling is 
$V_{\alpha \sigma' n', \beta \sigma n}  =  J_{\alpha \sigma' n', \beta \sigma n}
\; \Sv_{\alpha \sigma' n'} \cdot \Sv_{\beta \sigma n}$.
The first-order hopping matrix element in the one-triplon subspace is
\begin{equation}
\baln
t_{\alpha \sigma' n', \beta \sigma n}
&=
\left\langle n', \sigma', m \right|
V_{\alpha \sigma' n', \beta \sigma n}
\left|n, \sigma, m \right\rangle \\
&=
\frac{S(S+1)}{3}
J_{\alpha \sigma' n', \beta \sigma n}
\eta_\alpha\eta_\beta ,
\ealn
\label{eq:bond_matrix_element}
\end{equation}
with $\eta_A=+1$, $\eta_B=-1$, and
for $S=3/2$, $S(S+1)/3=5/4$.
Thus, like-site bonds (A-A) and (B-B) contribute with sign $+(5/4)J_{\alpha \beta}$, 
whereas unlike-site bonds (A-B) and (B-A) contribute with sign $-(5/4)J_{\alpha \beta}$.
For an illustration of the site and dimer labels and exchange couplings, see
Fig. \ref{fig:Js_MoBr3}.

Transforming to a Bloch sum basis, 
$\ket{\kv,\sigma, m} = \frac{1}{\sqrt{N}} \sum_n e^{i \kv \cdot \Rv_{n}} \ket{n, \sigma, m}$,
and  applying this rule to the intrachain exchange paths gives a one-dimensional triplon dispersion of the form
\begin{equation}
E_{\rm 1D}(k_y)
=
\Jc_1  - \tfrac{4}{5} D + A_y \cos(k_y),
\label{eq:one_d_triplon_general}
\end{equation}
where $k_y$ denotes the dimensionless wavevector $k_y b$ along the chain direction,
and $A_y  = - \frac{5}{2} \left( \Jc_2 - 2\Jc_3 + \Jc_{4} \right)$. 
The bottom of the one-dimensional triplon band occurs at $k_y=0$. 
The corresponding perturbative estimate of the singlet-triplet gap,
\be
\Delta_{\rm 1D}
=
\Jc_1  - \tfrac{4}{5} D - \frac{5}{2} \left( \Jc_2 - 2\Jc_3 + \Jc_{4} \right)
\label{eq:Delta_1D}
\ee
agrees closely with the independently computed DMRG gap, 
as shown in
Tables \ref{tab:MoBr3_params_vs_U}  and \ref{tab:MoI3_params_vs_U} of
Appendix \ref{app:Comp_methods}. 
The maximum percent difference is 5.2\% for \MoBr and  
2.0\% for \MoI.
This agreement provides a validation of the perturbative 
treatment of the interdimer
exchange coupling, and it demonstrates that the reduction of the gap from the 
isolated-dimer value is explained by coherent triplon hopping, 
and not by a breakdown of the singlet-dimer ground state.

\subsection{Three-dimensional triplon dispersion}

The full crystal contains two dimers per unit cell. 
The one-triplon Hamiltonian therefore becomes a $2\times2$ matrix in the dimer basis,
\begin{equation}
H_{\rm triplon}(\mathbf k)
=
\begin{pmatrix}
\epsilon(\mathbf k) & \gamma^*(\mathbf k)\\
\gamma(\mathbf k) & \epsilon(\mathbf k)
\end{pmatrix},
\label{eq:triplon_2x2}
\end{equation}
where symmetry-equivalent dimers have identical diagonal terms. 
The eigenvalues are
$E_\pm(\mathbf k)
=
\epsilon(\mathbf k)\pm |\gamma(\mathbf k)|$.
The diagonal term contains hopping between equivalent dimers,
\begin{equation}
\epsilon(\mathbf k)
=
\Jc_1  - \tfrac{4}{5} D
+
A_y\cos k_y
+
A_z\cos k_z
+
A_{yz}\cos k_y\cos k_z ,
\label{eq:epsilon_k}
\end{equation}
where 
$A_z=5J_5-5J_7$ and  $A_{yz} = -5 J_{10}$.
%
%
%
The magnitude of the off-diagonal coupling between dimers (I) and (II) is
%
%
\be
|\gamma (\kv) | = \cos(k_x/2) \cos(k_y/2) \sqrt{G_0^2 + G_z^2 + 2 G_0 G_z \cos(k_z)} 
\label{eq:abs_gamma}
\ee
with
$G_0  = -5J_4 + 10J_8 - 5J_{11}$ and
$G_z = -5J_6 + 10J_9$.
For \MoI, $G_0  = -5J_4 + 10J_8 - 5J_{12}$.
In Eqs. (\ref{eq:epsilon_k}) and (\ref{eq:abs_gamma}), $k_x$ and $k_z$
denote the dimensionless wavevector components
$k_x a$ and $k_z c$, respectively.
Note that the zero coupling along the zone-edge planes $k_x = \pi$ and 
$k_y=\pi$ that appears in Eq. (\ref{eq:abs_gamma}) through the factors $\cos(k_x/2) \cos(k_y/2)$
is a symmetry protected degeneracy resulting
from the screw-axis symmetries of the nonsymmorphic
\textit{Pmmn} space group. 
%
%
%

%
%
%
%
%
For the exchange constants listed in Tables 
\ref{tab:MoBr3_params_vs_U}  and  \ref{tab:MoI3_params_vs_U},
the minimum of the lower triplon branch occurs at Z.
The full three-dimensional excitation gap is then 
\be
\Delta_{\rm 3D}  \equiv  E_-(Z) 
= \Jc_1 - \tfrac{4}{5} D + A_y - A_z - A_{yz}  - 
\left| G_0 - G_z \right|.
\label{eq:Delta_3D}
\ee
The calculated $\Delta_{\rm 3D}$ is on the order of 1 meV below the single-chain gap, showing that 
interchain exchange is extremely weak and the physics is determined by the
strongly dimerized 1D chains.
Plots of the triplon dispersions along the high symmetry lines
are provided in the SM.
The persistence of a large positive gap is the central 
theoretical evidence for a robust singlet-dimer phase in the bulk crystal.

\subsection{INS Analysis and Discussion}
We use
a combination of powder averaged spin and phonon calculations to 
extract exchange, singlet-triplet gap, and triplon bandwidth values
from  our INS measurements 
described above.  
Phonon calculations were made using the INSPIRED software package \cite{HAN2024109288} and using the DFT determined force constants to calculate the powder average $S(Q,E)$ spectrum.  
These calculations were made for $T=0$ out to tenth order phonon modes.  
Calculated spectra were convolved with the calculated energy resolution in the region of the magnetic excitation and with the determined wave-vector resolution.  
The wave-vector resolution was determined from the width in wave-vector transfer of the first nuclear Bragg peak associated with the sample for elastically scattered neutrons.  
The Mo-Mo dimer bond of the MoBr$_3$ crystal structure corresponds to the real space fractional coordinate vector $\mathbf{d}_i=[0, 0.473, 0]$ ($|\mathbf{d}_i|=2.874$~\AA).  The inter-dimer Mo-Mo bond vector corresponds to $\mathbf{d}_i^{\prime}=[0, 0.527, 0]$ ($|\mathbf{d}_i^{\prime}|=3.203$~\AA), such that the dimer-dimer vector corresponds to $\mathbf{u}_0 = d_i + d_i^{\prime} = [0, 1, 0]$ ($|u_0| = 6.077$~\AA) with alternating chains along the crystallographic $b$-axis.  We calculate the dynamic structure factor of the alternating Heisenberg spin chain using
\begin{equation}
S(\mathbf{Q},\hbar\omega) = \mathcal{A}|f(Q)|^2 ( 1-\cos(\mathbf{Q}\cdot\mathbf{{d}_i}))\delta(\hbar\omega-\hbar\omega(\mathbf{Q}))
\end{equation}
where $\mathcal{A}$ is a multiplicative prefactor and $f(Q)$ is the neutron scattering magnetic form-factor of the Mo ion \cite{2000_coppernitrate}. 
$\hbar\omega(\mathbf{Q})$ in the above equation is the one-dimensional dispersion of the alternating chain model which we parameterize as
\begin{equation}
\hbar\omega(\mathbf{Q}) = \Jc_1  - \frac{4}{5}D+ A_y\cos(\mathbf{Q}\cdot\mathbf{u}_0),
\label{eq:hw(Q)}
\end{equation}
where $\Jc_1$, $D$, and $A_y$ are terms in the Heisenberg exchange Hamiltonian.  
We calculated the powder averaged and resolution convolved spectrum for $S(\mathbf{Q},\hbar\omega)$ as a function of $(\Jc_1  - \frac{4}{5}D)$ and $A_y$ \cite{2019AgVOAsO4}.
We found that a damped harmonic oscillator lineshape (Lorentzian) with a width of $\Gamma=1.4$~meV convolved with the data provided a more accurate representation of the data. 
This additional broadening beyond the instrumental resolution could represent finite spin lifetimes, a broader resolution function of the instrument configuration than what was calculated or additional exchange terms in the Hamiltonian that are not represented by the one-dimensional Hamiltonian \cite{zheludev2007reslib}.  
This result was linearly combined with the powder average phonon calculation and the relative phonon and spin contributions to the spectrum were fitted to a subset of the  measured spectrum as shown in Fig.~\ref{fig:neutron}(a).  
An overall constant background was included in this refinement.  
Best fit parameters correspond to $(\Jc_1  - \frac{4}{5}D) = 41.3 \pm0.8$~meV and $A_y = -10.8 \pm 0.7$~meV.  
The resulting total refined scattering function is shown in Fig.~\ref{fig:neutron}(b) with the spin-only contribution shown in Fig.~\ref{fig:neutron}(c) and the phonon only contribution shown in Fig.~\ref{fig:neutron}(d).  
These parameters correspond to a singlet-triplet gap value of $\Delta = 30.5 \pm 1.1$~meV in agreement with the thermodynamic measurements.  
Figure~\ref{fig:neutronIvsE} shows the calculated lineshape as a function of energy transfer through the wave-vector integrated scattering intensity.  
This lineshape is integrated over the identical measurement range of the calculation as is performed for the data.   
The lineshape was fitted to the wave-vector integrated scattering intensity with an overall multiplicative prefactor.

Comparing the refined parameters with the DFT values of Table \ref{tab:MoBr3_params_vs_U}
identifies where the 
first-principles description succeeds and where it does not.
The refined band center $\Jc_1 - \frac{4}{5}D = 41.3 \pm 0.8$ meV 
lies 12-17\% below 
the calculated values of 46.3 and 48.4 meV for $U_{\rm eff}=$ 1.1 and 1.2 eV.
The refined $A_y = -10.8 \pm 0.7$ meV is smaller than the calculated values
of -17.4 and -17.9 meV, so that the calculated triplon bandwidth $|2  A_y| =$ 
35-36 meV exceeds the measured $21.6 \pm 1.4$ meV value by 60\%. 
Because the gap $\Delta$ is given by the difference
$(\Jc_1 - \frac{4}{5} D) - |A_y|$,
these two overestimates cancel, and the 1D (3D) gaps calculated from perturbation
theory of 28.9 - 30.5 (28.4 - 30.1) meV agree with the measured value 
more closely than either quantity does individually.
Changing $U_{\rm eff}$ does not reconcile the differences, since $\Jc_1$ increases
more rapidly with $U_{\rm eff}$ than $|A_y|$, so that the gap $\Delta$ increases
with increasing $U_{\rm eff}$. 
We note that the use of the 1D dispersion in the refinement of Eq. (\ref{eq:hw(Q)})
is justified {\em a posteriori}. 
The exchange constants of Table \ref{tab:MoBr3_params_vs_U} 
give a total transverse bandwidth of $\sim 1.2$ meV,
below the instrumental resolution at 30 meV energy transfer and comparable
to the 1.4 meV broadening required to describe the lineshape.
The corresponding shift of the band minimum of 0.4 meV lies within
the uncertainty of the refined gap.

Although the numerical values of $\mathcal{J}_1$ and $A_y$ depend on $U_{\rm eff}$, 
every value considered, 
from 0 to 1.2~eV, yields the same hierarchy 
$\mathcal{J}_1 \gg |\mathcal{J}_n|, |J_m|$, 
a singlet DMRG ground state, and a finite singlet-triplet gap. 
Thus, 
the identification of MoBr$_3$ as a strongly dimerized, 
gapped spin-singlet 
system is insensitive to that choice. 
A classical ordered state would instead 
produce low-energy transverse spin waves, 
which are absent from the measured spectrum.

We also considered the alternative interpretation suggested by the reduced DFT local moment, 
namely that the effective magnetic degree of freedom on each Mo site is closer to $S=1$ 
than to $S=3/2$. 
In a classical energy mapping of a single exchange bond $H=J\mathbf S_1\cdot\mathbf S_2$, 
the energy difference between ferromagnetic and antiferromagnetic alignment is $\Delta E_{\rm FM-
AFM}=2JS^2$. 
Therefore, for the same DFT energy difference, the exchange constant extracted with 
$S=1$ is larger than that extracted with $S=3/2$ by a factor $9/4$. 
When these $S=1$
normalized exchange constants are used in the dimerized-chain calculation, 
the resulting singlet-triplet gaps are a factor of 
two larger than the $\Delta\approx 30$ meV gap 
measured by INS.
The reduced DFT local moment is therefore more naturally interpreted as a consequence 
of covalency and the spatial distribution of the spin density, rather than as evidence for an 
$S=1$ local spin model.

\section{Conclusion}
\label{sec:conclusion}
The XPS, thermally activated magnetic susceptibility,
fine resolution INS spectra,
and DFT / DMRG / perturbation theory analysis
all point to a consistent picture of an $S = 3/2$ quantum spin singlet 
ground state and a large singlet-triplet gap of $\sim 30$ 
meV for \MoBr. 
The INS data show the bandwidth of the \MoBr triplon band
to be approximately 20 meV. 
The DFT / perturbation theory analysis shows that the
triplon bandwidth is due to exchange coupling along the 1D chains
and that interchain coupling gives rise to negligible
additional bandwidth on the order of 1 meV.
The thermally activated magnetic susceptibility data and theoretical analysis
point to a nearly identical picture for \MoI
with an extracted singlet-triplet gap of $\sim 30$ meV.
These results provide the first experimental identification of a quantum spin-singlet ground state in a quasi-one-dimensional van der Waals material. 
Previous treatments of \MoX magnetic properties assumed classically ordered ground states with low-energy magnon excitations, 
whereas the present results demonstrate a singlet-dimer ground state, a large excitation gap, and a dispersive triplon band.

\section*{Data Availability}
The data supporting the findings of this study are available from the corresponding author upon reasonable request.

\begin{acknowledgements}
This work was supported in part by 
the Vannevar Bush Faculty Fellowship (VBFF) to A.A.B. from the 
Office of Naval Research (ONR) contract N00014-21-1-2947 on One-Dimensional Quantum Materials.
The NSF-MRI award 1625776 is acknowledged for the SQUID in the Material Characterization Facility of the UCLA MIC. 
The work at UCR and the University of Georgia was supported, in part, 
via subcontracts of the ONR project N00014-21-1-2947. 
F.K. and A.A.B. also acknowledge the support of the National Science Foundation (NSF), 
Division of Materials Research (DMR) via the project No. 2205973 entitled 
“Controlling Electron, Magnon, and Phonon States in Quasi‐2D Antiferromagnetic 
Semiconductors for Enabling Novel Device Functionalities.” C.M. acknowledges support through the Quantum Networks Training and Research Alliance in the Southeast (QuaNTRASE), funded by the National Science Foundation through award No. 2152159.  
DFT calculations were performed on STAMPEDE3 at TACC and EXPANSE at SDSC 
under allocation DMR130081 from the Advanced Cyberinfrastructure Coordination Ecosystem: 
Services $\&$ Support (ACCESS) program \cite{ACCESS}, 
which is supported by National Science Foundation grants $\#$2138259, $\#$2138286, $\#$2138307, $\#$2137603, and $\#$2138296.
DMRG calculations used UCR HPCC resources funded by grants from 
NSF (MRI-2215705, MRI-1429826) and NIH (1S10OD016290-01A1).
Magnetic force theorem calculations were performed on the Compute and Data Environment for Science (CADES) at the Oak Ridge National Laboratory, which is supported by the Office of Science of the U.S. Department of Energy under Contract No. DE-AC05-00OR22725.
This work was supported by the U.S. Department of Energy, Office of Science, Office of Basic Energy Sciences, Materials Sciences and Engineering Division (S.K. and M.Y.) and by the U.S. Department of Energy (DOE), Office of Science, National Quantum Information Science Research Centers, Quantum Science Center (D.H.K.).
We thank T. Oluwadara for assistance with powder XRD. We also thank Dr. M. Adams and Dr. W. Jin for providing access to the characterization 
facilities at the Alabama Micro/Nano Science and Technology Center (AMNSTC). 
H.G. is partially supported by the U.S. Department of Energy, Office of Science, under Grant No. DE-SC0023478.
This research used resources at the Spallation Neutron Source, a DOE Office of Science User Facility operated by the Oak Ridge National Laboratory. The beam time was allocated to SEQUOIA on proposal number IPTS-38410.1.
\end{acknowledgements}

\appendix

\section{Experimental Procedures and Measurement Details}
\label{app:Exp_methods}
\subsection{Crystal growth}
Single crystals of \MoI were grown by chemical vapor transport in a sealed fused-quartz ampule.
A 0.1022 g (0.705 mmol) portion of NH$_4$I powder (Fisher Scientific, 99.0\%) was placed at the bottom of a nitric acid-cleaned and dried ampule approximately $18 \times 2.6$ cm in size (22 mm inner diameter, 26 mm outer diameter, and $\sim 80$ cm$^3$ volume).
This was followed by 1.5073 g (11.877 mmol) of I$_2$ crystals (JT Baker, 99.9\%) and 0.3818 g (3.979 mmol) of Mo powder (Strem, 99.95\%).
The reagents were transferred in an Ar-filled glovebox using a glass funnel and an anti-static brush.
While submerged in an acetonitrile/dry ice bath, the ampule was evacuated and backfilled with Ar four times on a Schlenk line before being sealed under vacuum.
The ampule was placed in a horizontal tube furnace, and the temperature was ramped over 4 h to establish a gradient of 360 $^\circ$C in the source zone and 300 $^\circ$C in the growth zone.
After 240 h at this gradient, the ampule was cooled to room temperature over 6 h.
A total of 73.0 mg of lustrous silver, wire-like crystals was recovered from the growth zone (3.87\% isolated yield).
The crystals were stored in an Ar-filled glovebox.

Single crystals of \MoBr were grown in a sealed fused-quartz ampule under a temperature gradient.
Within an Ar-filled glovebox, 0.4524 g (4.715 mmol) of Mo powder (Strem, 99.95\%) was placed at the bottom of a pre-cleaned and dried ampule having a main body approximately 9 cm long (1.9 cm inner diameter and 2.2 cm outer diameter) and a 13 cm long neck (0.7 cm inner diameter and 1.0 cm outer diameter), with a total volume of $\sim 33$ cm$^3$.
A glass funnel and anti-static brush were used for clean transfer.
Next, 0.90 mL (17.6 mmol) of degassed Br$_2$ (Sigma-Aldrich, $\geq$99.5\%) was added by pipette.
While partially submerged in liquid nitrogen, the ampule was evacuated on a Schlenk line and sealed under vacuum.
The ampule was placed in a horizontal tube furnace, and the temperature was ramped over 6 h to establish a gradient of 350 $^\circ$C in the source zone and 300 $^\circ$C in the growth zone.
After 288 h at this gradient, the ampule was cooled to room temperature over 8 h.
A total of 1.4391 g of lustrous black \MoBr crystals was recovered from the growth zone (90.92\% isolated yield) and stored in an Ar-filled glovebox.

\subsection{Characterization}
The crystal morphology, elemental composition, and structure were characterized by 
SEM, EDS, and XRD.
SEM imaging was performed using an FEI Teneo FE-SEM at 10 keV with a spot size of 10, and EDS was performed using an Aztec Oxford Instruments X-MAX$^N$ detector operated at 10 keV.
For SEM and EDS analysis, as-grown crystals were mounted on a stub using carbon tape and mechanically exfoliated with Scotch tape.
The EDS maps show a homogeneous distribution of the constituent elements, with measured atomic ratios consistent with the stoichiometry of \MoX (X = I, Br).
XRD data were collected using a Bruker D2 Phaser diffractometer equipped with a LYNXEYE XE-T linear position-sensitive detector and Cu K$\alpha$ radiation ($\lambda = 1.5418$ \AA), operated at 30 kV and 10 mA.
Crystals were prepared as pressed mounts and rotated at 15 rotations min$^{-1}$, with a scan rate of 0.2 s step$^{-1}$.
Additional SEM images are provided in the Supplemental Material.

Surface chemical states were characterized by XPS using a Thermo Scientific K-Alpha X-Ray Photoelectron Spectrometer equipped with a monochromated Al K$\alpha$ x-ray source (1486 eV).
\MoI and \MoBr crystals were mounted on double-sided carbon tape adhered to a standard stage.
Preliminary spectra showed additional higher-binding-energy components attributable to surface oxidation (Fig.~S10).
To expose fresh surfaces, the crystals were mechanically exfoliated with 3M Scotch\textregistered{} Magic Tape immediately before loading.
The load lock was evacuated to a transfer pressure below $4.5 \times 10^{-7}$ mbar (approximately 45 min) before transfer to the analysis chamber.
Commercial MoO$_2$ (99\%) and MoO$_3$ ($\geq 99.5$\%) powders were placed in a powder sample holder and loaded similarly.
Survey spectra were collected with a pass energy of 200 eV, and high-resolution spectra were acquired with a pass energy of 50 eV and a spot size of 150 $\mu$m.

XPS binding-energy calibration and peak fitting were performed using a consistent procedure for all samples.
Binding energies (Table S1 in the Supplemental Material) were referenced to the C--C component of adventitious C 1s at 284.8 eV.
Spectra were processed and analyzed using Thermo Scientific Avantage software (Version 5.9931, Build 06755).
Background subtraction used the Smart background function, and peaks were fitted with mixed Gaussian--Lorentzian line shapes using the Powell fitting optimization algorithm.
The spin--orbit splitting and area ratios of the 3d$_{3/2}$--3d$_{5/2}$ doublets were constrained during fitting.

Reference XPS measurements on MoO$_2$ and MoO$_3$ were used to validate the energy calibration and spectral analysis.
These reference samples were measured under the same experimental conditions as the \MoI and \MoBr crystals.
Their measured binding energies and spectral line shapes (Fig.~S11) agree closely with established literature values for Mo(IV) and Mo(VI), respectively \cite{2010_MoO2&MoO3, 2015_MoOxide_XPS}.

Heat-capacity measurements were performed on \MoI and \MoBr using the thermal relaxation technique in a Quantum Design PPMS.
Because the naturally occurring \MoI crystals are thin, submillimeter strands that could not be mounted individually on the heat-capacity puck, the strands were gently rolled into a cotton ball-like aggregate exceeding the instrument's 1 mg minimum mass requirement and pressed into a compact pellet of $5.10 \pm 0.10$ mg.
For \MoBr, a single flake weighing $6.69 \pm 0.02$ mg was selected (Fig.~S13 in the Supplemental Material).
These preparation methods increased the sample contribution relative to the addenda and improved the signal-to-noise ratio and background subtraction.
Similar sample-mass optimization has been reported previously \cite{2023_Harry_Cp_method}.

Heat-capacity data were acquired from 1.8 to 100 K under high vacuum ($\approx 10^{-5}$ Torr) to ensure thermal isolation of the samples.
Before each run, the addenda contribution from Apiezon N grease was measured separately and subtracted from the total signal.
Each data point was recorded after thermal equilibration, and the relaxation curves were analyzed using the two-tau model implemented in the PPMS software.
Pointwise uncertainties from the two-tau fits were used as the primary error estimates.
Measurements were repeated twice for both compounds, with results reproducible within the instrument-reported uncertainties.
The typical uncertainty in $C_p$ was less than 5\% over the full temperature range.

Magnetic susceptibility was measured using a Quantum Design  MPMS3, 
which uses a superconducting quantum interference device (SQUID) with a 
sensitivity of $\leq 10^{-8}$ emu.
As-synthesized, randomly oriented \MoI (3.0 mg) and \MoBr (4.7 mg) crystals were attached with GE 7031 varnish to a quartz paddle holder (Fig.~S2 in the Supplemental Material).
The varnish and quartz paddle were selected for their small magnetic background.
ZFC and FC
susceptibility measurements were performed in vibrating sample magnetometry (VSM) mode from 1.8 to 300 K in an applied magnetic field of 0.5 T, using 2 K temperature increments.

Magnetic background subtraction and normalization were applied consistently to the susceptibility data.
Backgrounds were determined by identical measurements on an empty quartz paddle with a comparable amount of varnish and subtracted from the sample signals.
The susceptibilities were normalized by the applied magnetic field and the amount of sample in moles of formula units, following Ref.~\cite{2022_Mugiraneza_chi_tutorial}.
For additional verification reported in the Supplemental Material, two further \MoI samples (10 and 3.7 mg) and one further \MoBr sample (4.3 mg) were prepared and measured.

INS measurements were performed on approximately 1 g of \MoBr powder using the SEQUOIA chopper spectrometer at the Spallation Neutron Source \cite{SEQUOIA}.
The powder was sealed at room temperature in a cylindrical aluminum can 6.35 mm in diameter and 50 mm tall, with helium exchange gas to provide thermal contact.
The can was mounted in the bottom-loading sample changer of a closed-cycle helium refrigerator \cite{Stone2025}.
Measurements were made with incident neutron energy $E_i = 80$ meV, using the high-flux Fermi chopper at 180 Hz.
At each temperature, sample and empty-can measurements accumulated 11 and 5.5 C, respectively, of charge on the spallation target (approximately 4 and 2 h, respectively).

INS data were corrected for the empty-can background and detector response.
Empty-can data were acquired under the same conditions as the sample measurements, and vanadium standards were used to calibrate the detector efficiencies.
Neutron counts were normalized to the accumulated incident proton charge.
The data presented have the empty-can background subtracted, with uncertainties determined from Poisson counting statistics using conventional error propagation.

\section{Computational Details and Exchange Parameters}
\label{app:Comp_methods}
\subsection{DFT calculations}
DFT calculations were carried out using 
VASP \cite{kresse1993ab,kresse1996efficient}
based on the projector augmented wave (PAW) method \cite{kresse1999ultrasoft,blochl1994projector}
using the generalized gradient approximation
for the PBE exchange-correlation functional 
\cite{perdew1996generalized}.
The van der Waals
interaction is included using the PBE-D3 method of Grimme
et al. \cite{grimme2010consistent}.
Full structural relaxations were carried out using the conjugate-gradient algorithm until the residual Hellmann–Feynman forces on each atom were below $10^{-4}$ eV/\AA. 
Electronic self-consistency was achieved with a total-energy convergence criterion of $10^{-9}$ eV. 
A plane-wave energy cutoff of 520 eV and a suitably dense Monkhorst–Pack k-point mesh \cite{monkhorst1976special}
were employed to ensure energy convergence within 1 meV/atom. 
Typical k-point grids of $6\times12\times12$ were used for 
relaxation of the structures.
The relaxed lattice constants of
\MoI are $a=12.41$ \AA, $b=6.45$ \AA,
and $c = 7.17$ \AA,
and for \MoBr, they are
$a=11.47$ \AA, $b=6.06$ \AA, and
$c = 6.63$ \AA.
The second-order interatomic force constants are calculated using 
the supercell approach implemented in Phonopy \cite{Phonopy}.
For these calculations a $2\times2\times2$ supercell was used
with a k-point grid of $3\times6\times6$.

On-site Coulomb interactions for the transition-metal $d$ orbitals were treated within the DFT+$U$ formalism using the rotationally invariant Dudarev approach \cite{Dudarev}, in which only the effective parameter 
$U_{\mathrm{eff}} = U - J$ enters the Hamiltonian.
Calculations were performed for multiple values of $U_{\mathrm{eff}}$ to assess the sensitivity of the magnetic properties to electronic correlations.
For each chosen $U_{\mathrm{eff}}$, calculations were performed on the same crystal structure obtained from $U_{\mathrm{eff}}$=0 relaxation, with the atomic positions and lattice parameters kept fixed.
Previous first-principles studies of \MoBr and \MoI 
have used correlation treatments ranging from uncorrected PBE to PBE+U with Hubbard parameters of up to approximately 3 eV
with calculations on isolated dimerized chains using values between 
0.6 and 2 eV 
\cite{2017_MoX3_Kuzubov_JMMM,
2021_MoX3_ML_Zhang_PRB,
2024_MoI3_Spin_Chains_JMMM,
2025_Toroidic_PRB,
2025_Strain_Tunable_PRM}.
The values used for investigation of the exchange constants
described below lie comfortably within that range.

\subsection{Exchange interactions and magnetic anisotropy}
\begin{figure}[t]
\centering
\includegraphics[width=2.5in]{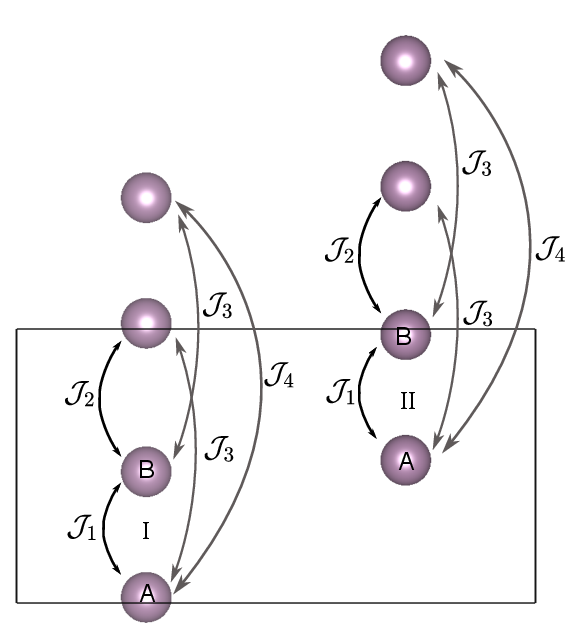}
\caption{
Supercell of \MoBr doubled along the chain direction showing only the Mo atoms.
Solid lines delineate the unit cell containing 4 Mo atoms forming
2 dimers parallel to the $b$ axis.
The two Mo atoms forming a dimer are labeled `$A$' and `$B$',
and the two dimers within each unit cell are labeled `I' and `II'.
The 4 nearest neighbor exchange couplings along the chains are shown.
}
\label{fig:Js_MoBr3}
\end{figure}

The isotropic Heisenberg exchange interactions were calculated using magnetic force linear response method \cite{liechtenstein1987local} implemented in the Jx package \cite{yoon2020jx}. 
To construct the Green’s function in a localized orbital basis, maximally localized Wannier functions were generated from the Mo $d$ and ligand (I and Br) $p$ orbitals using the Wannier90 package~\cite{pizzi2020wannier90}. 
The calculated exchange couplings were subsequently symmetrized according to the crystal symmetry using spglib~\cite{togo2024spglib}. 
The calculated exchange couplings were employed to construct the spin Hamiltonian as defined in Eq.~\ref{eq:heisenberg_general}.
The 4 intrachain exchange paths 
are illustrated in Fig. \ref{fig:Js_MoBr3}.
The convention of the Jx package is that a single 
dimer has Hamiltonian $H = 2J S_1 \cdot S_2$ and
$S_i$ is a normalized vector, so that the difference in
energy between the FM and AFM alignment is $4J$. 
Our convention 
is that $H = J S_1 \cdot S_2$, and the difference between
the FM and AFM alignments is $2JS^2$ with $S=3/2$. 
Thus, the exchange parameters from Jx are multiplied
by $2/S^2$ for use in our DMRG calculations.
Magnetocrystalline anisotropy was evaluated by including spin–orbit coupling (SOC) in non-collinear DFT calculations, with the anisotropy energies obtained from total-energy differences associated with magnetization oriented along different crystallographic axes.

For each material, exchange and anisotropy values calculated 
for two representative $U_{\rm eff}$ values are shown in
Tables \ref{tab:MoBr3_params_vs_U} and \ref{tab:MoI3_params_vs_U}.
The singlet-triplet gap resulting from these parameters
matches or brackets the measured values of $\sim 30$ meV.
As $U_{\rm eff}$ increases from zero, the largest effect is on the
value of $\mathcal{J}_1$, which monotonically increases.
The bandwidth resulting from interdimer coupling, $|2  A_y|$, increases slightly, but is much less sensitive
to the value of $U_{\rm eff}$.
Thus, the singlet-triplet gap increases with $U_{\rm eff}$.
\begin{table}
\centering
\caption{
\MoBr Exchange constants (meV) and single-ion anisotropy values (meV) for two different 
$U_{\rm eff}$ (eV) values. The exchange constants are ordered by distance.
Excitation gaps $\Delta_{1D}$ and $\Delta_{3D}$ are determined from 
Eqs. (\ref{eq:Delta_1D}) and (\ref{eq:Delta_3D}), respectively.
$\Delta_{\rm DMRG}$ and $\Delta_{\rm DMRG}^D$
are from DMRG calculations of single chains without and with anisotropy $D$,
respectively.
}
\begin{tabular}{|c|c|c|}
\hline
$U_{\rm eff}$  & 1.1  &  1.2 \\
\hline
\hline
$J_1(\Jc_1)$    & 46.664 & 48.763 \\
\hline
$J_2(\Jc_2)$    & 1.384 & 1.427 \\
\hline
$J_3(\Jc_3)$    & -1.936 & -2.009 \\
\hline
$J_4$           & 0.114 & 0.108 \\
\hline
$J_5$           & 0.124 & 0.118 \\
\hline
$J_6$           & 0.169 & 0.163 \\
\hline
$J_7$           & 0.133 & 0.130 \\
\hline
$J_8$           & 0.062 & 0.060 \\
\hline
$J_9$           & 0.057 & 0.055 \\
\hline
$J_{10}$        & 0.005  &  0.004 \\
\hline
$J_{11}$        & -0.035 & -0.035 \\
\hline
$J_{12}(\Jc_4)$ & 1.718  &  1.717 \\
\hline
$D$             & 0.459 & 0.436 \\
\hline
$\Delta_{1D}$   & 28.86 & 30.51 \\
\hline
$\Delta_{3D}$   & 28.43  & 30.09 \\
\hline
$\Delta_{\rm DMRG}$    & 30.91 & 32.56 \\
\hline
$\Delta_{\rm DMRG}^D$   & 30.43 & 32.11 \\
\hline
\end{tabular}
\label{tab:MoBr3_params_vs_U}
\end{table}
\begin{table}
\centering
\caption{
\MoI Exchange constants (meV) and single-ion anisotropy values (meV) for two different 
$U_{\rm eff}$ (eV) values. The exchange constants are ordered by distance.
Excitation gaps $\Delta_{1D}$ and $\Delta_{3D}$ are determined from 
Eqs. (\ref{eq:Delta_1D}) and (\ref{eq:Delta_3D}), respectively.
$\Delta_{\rm DMRG}$ and $\Delta_{\rm DMRG}^D$
are from DMRG calculations of single chains without and with anisotropy $D$,
respectively.
}
\begin{tabular}{|c|c|c|c|}
\hline
$U_{\rm eff}$   & 0.6 & 1.1 \\
\hline
\hline
$J_1(\Jc_1)$    & 32.76 & 41.884 \\
\hline
$J_2(\Jc_2)$    & 0.554 & -0.017 \\
\hline
$J_3(\Jc_3)$    & -1.173 & -1.263 \\
\hline
$J_4$           & 0.157 & 0.109 \\
\hline
$J_5$           & 0.218 & 0.159 \\
\hline
$J_6$           & 0.366  & 0.310 \\
\hline
$J_7$           & 0.270 & 0.231 \\
\hline
$J_8$           & 0.072 & 0.056 \\
\hline
$J_9$           & 0.053 & 0.026 \\
\hline
$J_{10}$        & -0.049 & -0.059 \\
\hline
$J_{11}(\Jc_4)$ & 0.471 & 0.442 \\
\hline
$J_{12}$        & -0.053 & -0.052 \\
\hline
$D$             & 0.722 &  0.569 \\
\hline
$\Delta_{1D}$   & 23.75  & 34.05 \\
\hline
$\Delta_{3D}$   & 22.27  & 32.56 \\
\hline
$\Delta_{\rm DMRG}$    & 24.87 & 34.82 \\
\hline
$\Delta_{\rm DMRG}^D$   & 24.23 & 34.31\\
\hline
\end{tabular}
\label{tab:MoI3_params_vs_U}
\end{table}
\subsection{DMRG calculations and finite-size extrapolation}
The DMRG calculations of single $S=3/2$ spin chains used open boundary conditions as implemented in TeNPy \cite{TeNPy}. 
Exchange couplings up to fourth nearest neighbor along the chain, $\Jc_1$, $\Jc_2$, $\Jc_3$, and $\Jc_{4}$, were included in the calculations.
The explicit, single-chain Hamiltonian is
\be
\begin{aligned}
H_N={}&J_1\sum_n \mathbf S_{A,n}\cdot\mathbf S_{B,n}
+J_2\sum_n \mathbf S_{B,n}\cdot\mathbf S_{A,n+1}\\
&+J_3\sum_n\left(
\mathbf S_{A,n}\cdot\mathbf S_{A,n+1}
+\mathbf S_{B,n}\cdot\mathbf S_{B,n+1}\right)\\
&+J_4\sum_n\mathbf S_{A,n}\cdot\mathbf S_{B,n+1}
+D\sum_i(S_i^z)^2 ,
\end{aligned}
\label{eq:H_single-chain}
\ee
where $n$ labels the unit cells, or equivalently, the
dimers.
The singular value cutoff was set
to $10^{-12}$, the energy-convergence was set
to $10^{-10}$, and the matrix-product-state norm 
tolerance was set to $10^{-6}$. 
The maximum bond dimension was set to 3000, but that
value was never reached.
The maximum bond dimension increased for longer chains
and was larger for the excited states compared to the
ground state.
A representative calculation for \MoBr with
$N=80$, and values for $\Jc_1$ - $\Jc_4$ and
$D$ given by Table \ref{tab:MoBr3_params_vs_U} for
$U_{\rm eff} = 1.1$ results in maximum
bond dimensions of 503 and 2209 for the
calculations of the $\Stot^z=0$ ground state and
first excited state, respectively, and
1597 for the calculation of the $\Stot^z=1$ 
lowest-energy state.

The thermodynamic-limit gap was obtained by extrapolating $\Delta_N$ 
as an even 4th order polynomial in $1/N$,
where $N$ is the even number of atoms in the chain. 
The even power form is motivated by the small
wavevector expansion of the cosine-like triplon
dispersion, together with the finite-chain
quantization $k_{\rm min} = \pi/L \propto 1/N$.
An example plot of the extrapolation is provided in
the SM.

In the isotropic limit, the ground state has $\Stot = 0$,
and the three degenerate lowest excited states form an
$\Stot = 1$ triplet.
In this limit,
the singlet-triplet excitation gap is obtained from
the ground state and first excited state of the $S_{\rm tot}^z=0$ block,
\begin{equation}
\Delta_N = E_1(N,S^z_{\rm tot}=0) - E_0(N,S^z_{\rm tot}=0),
\label{eq:finite_size_gap}
\end{equation}
where $E_0(N,S^z_{\rm tot}=0)$ is the singlet-sector ground-state energy, 
and $E_1(N,S^z_{\rm tot}=0)$ is the $S^z_{\rm tot}=0$ state of the triplet sector. 
The single ion anisotropy preserves $\Stot^z$ but not $\Stot^2$
and therefore splits the triplet into an $\Stot^z=0$ component and
a degenerate $\Stot^z = \pm 1$ pair.
Because the anisotropy is small and no level crossing occurs, these 
three states remain adiabatically connected to the isotropic
$\Stot=1$ triplet.
With the inclusion of a single-ion anisotropy term, the lowest energy of
the $S_{\rm tot}^z=1$ block is also calculated, and the gap is
determined from ${\rm min}[ E_1(N,S^z_{\rm tot}=0),\:
E_0(N,S^z_{\rm tot}=1)] - E_0(N,S^z_{\rm tot}=0)$. 
In all cases, in the presence
of the easy-plane anisotropy term, the lowest excited state energy is
$E_0(N,S^z_{\rm tot}=1)$ corresponding to the
$\Stot^z=\pm 1$ pair.

\section{Susceptibility of an $S=3/2$ Dimer}
\label{app:chi_Cv}
With a weak magnetic field, the Hamiltonian of Eq. (\ref{eq:Hdimer}) becomes
$H = H_{\rm dimer} - g \mu_B B \Stot^z$, and within an $S_{\rm tot}$ multiplet,
$E(\Stot,m) = E(\Stot) - g \mu_B B m$.
The partition function is 
$Z(B) = \sum_{\Stot = 0}^3 \;\; \sum_{m = -\Stot}^{\Stot}
\exp \left\{ -\beta \left[ E(\Stot) - g \mu_B B m \right] \right\}$, 
so that the magnetization is
$M = g \mu_B \left\langle \Stot^z \right\rangle = \frac{1}{\beta} \frac{\partial \ln Z}{\partial B}$,
and the susceptibility to the magnetic flux density is
\be
\baln
\chi_B =& \left. \frac{\partial M}{\partial B} \right|_{B=0}
=
\left. \frac{1}{\beta} \frac{\partial^2 \ln Z}{\partial B^2} \right|_{B=0} \\
= & - \left. \frac{1}{\beta Z^2}\left( \frac{\partial Z}{\partial B} \right)^2 \right|_{B=0}
+ \left. \frac{1}{\beta Z} \frac{\partial^2 Z}{\partial B^2} \right|_{B=0}
\ealn
\label{eq:d2ZdB2}
\ee
The first term is zero, since
\be
\left. 
\frac{1}{Z} \frac{\partial Z}{\partial B} \right|_{B=0} = 
\frac{\beta g \mu_B}{Z(B=0)} \sum_{\Stot = 0}^3 \exp \left\{ -\beta  E(\Stot)  \right\} 
\sum_{m = -\Stot}^{\Stot}  m
= 0 .
\ee
The second term of Eq. (\ref{eq:d2ZdB2}) gives
\be
\baln
\chi_B &= 
\frac{\beta\left( g \mu_B \right)^2}{Z(B=0)} 
\sum_{\Stot = 0}^3 \exp \left\{ -\beta  E(\Stot)  \right\}
\sum_{m = -\Stot}^{\Stot} m^2\\
&=
\beta \left( g \mu_B \right)^2 \left. \left\langle \left( \Stot^z \right)^2 
\right\rangle \right|_{B=0}
\ealn
\label{eq:chiB}
\ee
Using $\sum_{m=-\Stot}^{\Stot} m^2 = \Stot (\Stot + 1) (2 \Stot + 1)/3$,
we have,
\be
\chi_B = \beta \left( g \mu_B \right)^2 
\frac{\sum_{\Stot} e^{-\beta E(\Stot)}\Stot(\Stot+1)(2\Stot +1)/3}{\sum_{\Stot}  e^{-\beta E(\Stot)} (2\Stot + 1) }
\label{eq:chi_Stot}
\ee
Writing out each term of the sums in Eq. (\ref{eq:chi_Stot}) gives
\be
\chi_B = \frac{\left(g \mu_B \right)^2 }{k_B T}
\frac{2e^{-J/k_BT} + 10 e^{-3J/k_BT} + 28 e^{-6J/k_BT} }{1 + 3e^{-J/k_BT} + 5e^{-3J/k_BT} + 7e^{-6J/k_BT}}
\label{eq:chi_B_dimer}
\ee
where $J \equiv \Jc_1$ is the dimer exchange.
This is the susceptibility of a dimer (2 spins). 
For the susceptibility per mole of spins, multiply by 
Avogadro's number and divide by two.
This is the susceptibility in Gaussian units
($\chi_H = \chi_B$).
The $1/T$ prefactor is the form used in Eq. (\ref{eq:chi_t2}).

\section{Effect of dimensionality on the temperature prefactor of the susceptibility}
\label{sec:T-prefactor}
We now analyze the effect of dimensionality and triplon bandwidth on the 
power of the temperature prefactor of $\chi$.
Considering only the triplet excitations, 
we can write the susceptibility as
\be
\chi(T) \approx n(T) \frac{(g \mu_B)^2}{k_BT} ,
\label{eq:chi-T-approx}
\ee
where $n(T)$ is the number density of triplets per unit length.
For the infinite chain, the triplons form a band with a dispersion
$E(k) = J_1 - \frac{W}{2} \cos(k)$, where $W$ is the bandwidth,
so that
$n(T) \approx \int_{-\pi}^\pi \frac{dk}{2\pi} e^{-\beta E(k)} 
= e^{-\beta J_1} I_0\left( \frac{\beta W}{2} \right)$,
where $I_0$ is the modified Bessel function of the first kind.
We consider two cases.
In the low-temperature limit when the temperature is much less than the
bandwidth, $(\beta W \gg 1)$, we use the asymptotic
form 
$I_0(x) \approx \frac{e^x}{\sqrt{2\pi x}}$
so  that 
$n(T) \sim \sqrt{\frac{k_BT}{W}} e^{-\beta \Delta}$, where $\Delta = J_1 - W/2$.
Placing this form back in Eq. (\ref{eq:chi-T-approx}) gives the 1D prefactor
of $1/\sqrt{T}$.
In the high-temperature limit
when the temperature is much greater than the bandwidth, $(\beta W \ll 1)$,
$I_0(x) \approx 1 $.
Therefore,
$\int \frac{dk}{2\pi} e^{-\beta E(k)} \approx e^{-\beta J_1}$, where
$J_1$ is the center of the triplon band.
Thus, for $k_BT \gg W$, the dispersion is thermally smeared
and the integral approaches a constant value $n(T) \sim e^{-\beta J_1}$.
Placing this form back into Eq. (\ref{eq:chi-T-approx}) gives the 0D,
single-dimer prefactor of $1/T$.

If we now assume a separable interchain dispersion along $\av$ and $\cv$
such that
$E(\kv) = J_1 - \frac{W_b}{2} \cos(k_b) - \frac{W_a}{2}\cos(k_a) - \frac{W_c}{2}\cos(k_c)$,
then each transverse dimension contributes a term $I_0(\beta W_j/2)$,
which in the low temperature limit gives a factor of $T^{1/2}$. 
Thus, the low-temperature temperature coefficient becomes
$T^{d/2-1}$, where $d$ is the dimensionality.
In the high temperature limit where $\beta  W_j \ll 1$, each dimension $j$
contributes a factor of $\sim 1$.
If the dispersion along the chain, $W_b$, is much
greater than the dispersion transverse to the chains, $W_b \gg W_t = W_a + W_c$,
then, when $k_B T$ is much greater than the transverse band width, 
$k_B T \gg W_t$, the temperature dependence of $\chi$
reverts back to the 1D form $T^{-1/2}$.
As $k_B T$ increases and finally becomes greater than the full bandwidth, 
$W=W_a+W_b+W_c$, the temperature coefficient transitions to the 0D (isolated dimer)
value of  $T^{-1}$.

The INS data show a bandwidth of the triplon band of 
approximately 20 meV.
The temperature window $k_BT$ becomes greater than the triplon bandwidth before
the peak in the susceptibility curves, placing the experimental responses in the crossover between the one-dimensional and isolated-dimer limiting forms.

\clearpage
\newpage
\bibliography{RL.bib}

@PREAMBLE{
 "\providecommand{\noopsort}[1]{}" 
 # "\providecommand{\singleletter}[1]{#1}%" 
}

@article{TeNPy,
title = {Efficient numerical simulations with Tensor Networks: Tensor Network {Python} ({TeNPy})},
author = {J. Hauschild and F. Pollmann},
journal = {SciPost Physics Lecture Notes},
volume = {5},
pages = {1--32},
year = {2018}
}

@article{2017_MoX3_Kuzubov_JMMM,
title = {DFT investigation of electronic structures and magnetic properties of halides family {MeHal}$_3$ ({Me = Ti, Mo, Zr, Nb, Ru, Hal = Cl, Br, I}) one dimensional structures},
journal = {Journal of Magnetism and Magnetic Materials},
volume = {440},
pages = {93-96},
year = {2017},
note = {Selected papers from the sixth Euro-Asian Symposium “Trends in Magnetism” (EASTMAG-2016)},
issn = {0304-8853},
doi = {10.1016/j.jmmm.2016.12.054},
url = {https://doi.org/10.1016/j.jmmm.2016.12.054},
author =  {A. A. Kuzubov and E. A. Kovaleva and M. I. Popova and A. S. Kholtobina and N. S. Mikhaleva and M. A. Visotin and A. S. Fedorov}
}

@article{2021_MoX3_ML_Zhang_PRB,
  title = {Bipolar ferromagnetic semiconductors and doping-tuned room-temperature half-metallicity in monolayer {$\mathrm{MoX}_{3}$ $(X = \mathrm{Cl}, \mathrm{Br}, \mathrm{I})$: an HSE06} study},
  author = {Zhang, Jiayong and Zhao, Bao and Ma, Chunlan and Yang, Zhongqin},
  journal = {Phys. Rev. B},
  volume = {103},
  issue = {7},
  pages = {075433},
  numpages = {16},
  year = {2021},
  month = {Feb},
  publisher = {American Physical Society},
  doi = {10.1103/PhysRevB.103.075433},
  url = {https://doi.org/10.1103/PhysRevB.103.075433}
}

@article{1972_MoBr3_Structure,
title = {{Die Verfeinerung der {MoBr$_3$}-Struktur}},
journal = {Journal of Solid State Chemistry},
volume = {4},
number = {3},
pages = {410-416},
year = {1972},
issn = {0022-4596},
doi = {https://doi.org/10.1016/0022-4596(72)90156-9},
url = {https://www.sciencedirect.com/science/article/pii/0022459672901569},
author = {Dietrich Babel}
}

@article{2022_MoI3_Fari_APL,
author = {Kargar,Fariborz  and Barani,Zahra  and Sesing,Nicholas R.  and Mai,Thuc T.  and Debnath,Topojit  and Zhang,Huairuo  and Liu,Yuhang  and Zhu,Yanbing  and Ghosh,Subhajit  and Biacchi,Adam J.  and da Jornada,Felipe H.  and Bartels,Ludwig  and Adel,Tehseen  and Hight Walker,Angela R.  and Davydov,Albert V.  and Salguero,Tina T.  and Lake,Roger K.  and Balandin,Alexander A. },
title = {Elemental excitations in {MoI$_3$} one-dimensional van der {Waals} nanowires},
journal = {Applied Physics Letters},
volume = {121},
number = {22},
pages = {221901},
year = {2022},
doi = {10.1063/5.0129904},
URL = {https://doi.org/10.1063/5.0129904}
}

@article{2024_Topo_Mag_AFM_Chain_PRB,
  title = {Topological magnonic properties of an antiferromagnetic chain},
  author = {Debnath, Topojit and Soundararaj, Shri Hari and Kwon, Sohee and Balandin, Alexander A. and Lake, Roger K.},
  journal = {Phys. Rev. B},
  volume = {110},
  issue = {14},
  pages = {144448},
  numpages = {14},
  year = {2024},
  month = {Oct},
  publisher = {American Physical Society},
  doi = {10.1103/PhysRevB.110.144448},
  url = {https://doi.org/10.1103/PhysRevB.110.144448}
}

@article{2025_Toroidic_PRB,
  title = {Tunable toroidic phase transition in one-dimensional antiferromagnetic ferrotoroidic semiconductor},
  author = {Zou, Jin and Li, Wuxuan and Xun, Yechen and Wang, Hua and Qi, Jingshan},
  journal = {Phys. Rev. B},
  volume = {111},
  issue = {13},
  pages = {134416},
  numpages = {11},
  year = {2025},
  month = {Apr},
  publisher = {American Physical Society},
  doi = {10.1103/PhysRevB.111.134416},
  url = {https://link.aps.org/doi/10.1103/PhysRevB.111.134416}
}

@article{2025_Strain_Tunable_PRM,
  title = {Strain-tunable and carrier-driven structural and magnetic phase transitions in one-dimensional dimerized $\mathrm{Mo}{X}_{3}$ {$(X=\mathrm{Br},\mathrm{I})$} atomic chains},
  author = {Ma, Xinru and Lei, Yuchen and Bao, Hongwei and Wu, Wenting and Ma, Fei and Li, Yan},
  journal = {Phys. Rev. Mater.},
  volume = {9},
  issue = {3},
  pages = {034412},
  numpages = {14},
  year = {2025},
  month = {Mar},
  publisher = {American Physical Society},
  doi = {10.1103/PhysRevMaterials.9.034412},
  url = {https://link.aps.org/doi/10.1103/PhysRevMaterials.9.034412}
}

@article{2004_RuBr3_MoBr3_ChemMat,
author = {Merlino, Stefano and Labella, Luca and Marchetti, Fabio and Toscani, Simone},
title = {Order-Disorder Transformation in {RuBr$_3$} and {MoBr$_3$}: A Two-Dimensional {Ising} Model},
journal = {Chemistry of Materials},
volume = {16},
number = {20},
pages = {3895-3903},
year = {2004},
doi = {10.1021/cm049235q},
URL = {https://doi.org/10.1021/cm049235q}
}

@article{2020_MoI3_photodetectors,
  title = {Low ligand field strength ion {(I-)} mediated {1D} inorganic material {MoI$_3$}: Synthesis and application to photo-detectors},
  author = {Kyung Hwan Choi and Seungbae Oh and Sudong Chae and Byung Joo Jeong and Bum Jun Kim and Jiho Jeon and Sang Hoon Lee and Sang Ok Yoon and Chaeheon Woo and Xue Dong and Asghar Ghulam and Changmo Lim and Zhixiang Liu and Cong Wang and Ali Junaid and Jae-Hyun Lee and Hak Ki Yu and Jae-Young Choi},
  journal = {Journal of Alloys and Compounds},
  volume = {853},
  pages = {157375},
  numpages = {14},
  year = {2021},
  month = {Feb},
  doi = {10.1016/j.jallcom.2020.157375},
  url = {https://doi.org/10.1016/j.jallcom.2020.157375}
}

@article{2026_MoI3_pellet,
  title = {A pressed bulk pellet {MoI$_3$} photodetector with high photocurrent},
  author = {Lifang Wu and Xueqing Wang and Chang Xue and Tianhao Wu and Ihsan Ullah and Selcuk Terci and Shenghao Wang},
  journal = {AIP Advances},
  volume = {16},
  pages = {065018},
  numpages = {6},
  year = {2026},
  month = {June},
  doi = {10.1063/5.0337096},
  url = {https://doi.org/10.1063/5.0337096}
}

@article{2023_Mo6Br12,
  title = {Unusual Square Pyramidal Chalcogenide {Mo$_5$} Cluster with Bridging Pyrazolate-Ligands},
  author = {Lulina V Savina and Anton A Ivanov and Darya V Evtushok and Yakov M Gayfulin and Andrey Y Komarovskikh and Mikhail M Syrokvashin and Mariia N Ivanova and Igor P Asanov and Ilia V Eltsov and NAtalia V Kuratieva and Yuri V Mironov and Michael A Shestopalov},
  journal = {Int. J. Mol. Sci},
  volume = {24},
  pages = {3440},
  numpages = {4},
  year = {2023},
  month = {June},
  doi = {10.3390/ijms24043440},
  url = { https://doi.org/10.3390/ijms24043440}
}

@article{1974_Mo6Cl12,
  title = {X-ray photoelectron spectra of inorganic molecules. {IX}. Distinction between bridging and terminal metal-chlorine bonds in metal halide clusters of rhenium(III) and molybdenum(II)},
  author = {A D Hamer and R A Walton},
  journal = {Inorg. Chem},
  volume = {13},
  pages = {1446},
  numpages = {5},
  year = {1974},
  month = {June},
  doi = {10.1021/ic50136a038},
  url = { https://doi.org/10.1021/ic50136a038}
}

@article{1978_Mo6I12,
  title = {Studies on metal carboxylates. 13. A high-yield synthesis of rhenium(III) iodide and the isolation and characterization of $\beta$-molybdenum(II) iodide.},
  author = {H. D. Glicksman and R. A. Walton},
  journal = {Inorg. Chem},
  volume = {17},
  number = {1},
  pages = {200-201},
  numpages = {2},
  year = {1978},
  month = {January},
  doi = {10.1021/ic50179a040},
  url = {https://doi.org/10.1021/ic50179a040}
}

@article{2010_MoO2&MoO3,
  title = {Theoretical and Experimental Study of the Electronic Structures of {MoO$_3$} and {MoO$_2$}},
  author = {David O Scanlon and Graeme W Watson and D J Payne and G R Atkinson and R G Egdell and D S L Law},
  journal = {J. Phys. Chem. C},
  volume = {114},
  number = {10},
  pages = {4636-4645},
  numpages = {9},
  year = {2010},
  month = {March},
  doi = {10.1021/jp9093172},
  url = { https://doi.org/10.1021/jp9093172}
}

@article{2015_MoOxide_XPS,
  title = {Generalized molybdenum oxide surface chemical state {XPS} determination via informed amorphous sample model.},
  author = {Jonas Baltrusaitis and Beatriz Mendoza-Sanchez and Vincent Fernandez and Rick Veenstra and Nijole Dukstiene and Adam Roberts and Neal Fairley},
  journal = {Applied Surface Science},
  volume = {326},
  pages = {151-161},
  numpages = {10},
  year = {2015},
  month = {},
  doi = {10.1016/j.apsusc.2014.11.077},
  url = { https://doi.org/10.1016/j.apsusc.2014.11.077}
}

@article{Gruber1983,
  author  = {H. Gruber and H. Haselmaier and H.P. Fritzer},
  title   = {On Magnetic Properties of Some Molybdenum Oxides},
  journal = {Journal of Solid State Chemistry},
  year    = {1983},
  volume  = {47},
  pages   = {84-91},
  doi     = {10.1016/0022-4596(83)90045-2},
  url     = {https://doi.org/10.1016/0022-4596(83)90045-2}
}

@article{Sharmistha_2024,
  author  = {Sharmistha Dey and Vikash Mishra and Neetesh Dhakar and Sunil Kumar and Pankaj Srivastava and Santanu Ghosh},
  title   = {Probing defect induced room temperature ferromagnetism in Chemical Vapor Deposition grown {MoO$_3$} flakes: A correlation with electronic structure and first principle-based calculations},
  journal = {Journal of Alloys and Compounds},
  year    = {2024},
  volume  = {980},
  pages   = {173650},
  doi     = {10.1016/j.jmst.2014.08.013},
  url     = {https://doi.org/10.1016/j.jmst.2014.08.013}
}

@article{Sandeep_2015,
  author  = {Sandeep K.S. Patel and Khemchand Dewangan and N.S. Gajbhiye},
  title   = {Synthesis and Room Temperature d$^0$ Ferromagnetic Properties of $\alpha$-{MoO$_3$} Nanofibers},
  journal = {Journal of Materials Science \& Technology},
  year    = {2015},
  volume  = {31},
  number  = {5},
  pages   = {453-457},
  doi     = {10.1016/j.jmst.2014.08.013},
  url     = {https://doi.org/10.1016/j.jmst.2014.08.013}
}

@article{Schlenker_1986,
  author  = {C. Schlenker and S.S.P. Parkin and H. Guyot},
  title   = {Landau diamagnetism in the charge density wave state of $\gamma$-{Mo$_4$O$_{11}$}},
  journal = {Journal of Magnetism and Magnetic Materials},
  year    = {1986},
  volume  = {54-57},
  number  = {3},
  pages   = {1313-1314},
  doi     = {10.1016/0304-8853(86)90834-6},
  url     = {https://doi.org/10.1016/0304-8853(86)90834-6}
}

@article{MZoller_2020,
  author  = {Michael Zoller and Rimma Bubnova and Yaroslav Biryukov and Eiken Haussühl and Rainer Pöttgen and
  Oliver Janka and Simon Penner and Corsin Praty and Harald Fitzek and Jörg Winkler and Stanislav Filatov
  and Hubert Huppertz},
  title   = {Elucidating the physical properties of the
  molybdenum oxide {Mo$_4$O$_{11}$} and its tantalum
  substituted variant {Mo$_2$Ta$_2$O$_{11}$}},
  journal = {Zeitschrift für Kristallographie - Crystalline Materials},
  year    = {2020},
  volume  = {235},
  number  = {4-5},
  pages   = {143-155},
  doi     = {10.1515/zkri-2019-0073},
  url     = {https://doi.org/10.1515/zkri-2019-0073}
}

@article{Nomura_2017,
  title = {$\ensuremath{\alpha}\ensuremath{-}\ensuremath{\beta}$ and $\ensuremath{\beta}\ensuremath{-}\ensuremath{\gamma}$ phase boundaries of solid oxygen observed by adiabatic magnetocaloric effect},
  author = {Nomura, T. and Kohama, Y. and Matsuda, Y. H. and Kindo, K. and Kobayashi, T. C.},
  journal = {Phys. Rev. B},
  volume = {95},
  issue = {10},
  pages = {104420},
  numpages = {6},
  year = {2017},
  month = {Mar},
  publisher = {American Physical Society},
  doi = {10.1103/PhysRevB.95.104420},
  url = {https://link.aps.org/doi/10.1103/PhysRevB.95.104420}
}

@ARTICLE{7405275,
  author={Ramos-Álvarez, A. and Sóñora, D. and Dancausa, J. D. and Doval, J. M. and Verde, J. C. and Mosqueira, J. and Vidal, F.},
  journal={IEEE Transactions on Applied Superconductivity}, 
  title={A Procedure to Protect Granular and Porous Samples From Oxygen Contamination in High-Resolution Magnetization Measurements: Application to Granular High-{$T_{c}$} Cuprate Superconductors}, 
  year={2016},
  volume={26},
  number={3},
  pages={1-3},
  doi={10.1109/TASC.2016.2529427}}

@article{2023_Harry_Cp_method, 
  author={Goyal, Harshil and Peek, Sherman E. and Sellers, John A. and Hamilton, Michael C.},
  journal={IEEE Transactions on Applied Superconductivity}, 
  title={Methodology to Characterize Thermal Properties of Thin Film Superconductors Using a DynaCool Physical Property Measurement System}, 
  year={2023},
  volume={33},
  number={5},
  pages={1-5},
  doi={10.1109/TASC.2023.3271950}
}

@article{2000_Thermo_S1-2_AFM_Chains_PRB,
  title = {Thermodynamics of spin {$S=1/2$} antiferromagnetic uniform and alternating-exchange {Heisenberg} chains},
  author = {Johnston, D. C. and Kremer, R. K. and Troyer, M. and Wang, X. and Kl\"umper, A. and Bud'ko, S. L. and Panchula, A. F. and Canfield, P. C.},
  journal = {Phys. Rev. B},
  volume = {61},
  issue = {14},
  pages = {9558--9606},
  numpages = {0},
  year = {2000},
  month = {Apr},
  publisher = {American Physical Society},
  doi = {10.1103/PhysRevB.61.9558},
  url = {https://link.aps.org/doi/10.1103/PhysRevB.61.9558}
}

@article{2011_Spin_Dyn_Organic_Lin_Chains_PRB,
  title = {Spin dynamics of the organic linear chain compounds {(TMTTF)${}_{2}X$ ($X$ = SbF${}_{6}$, AsF${}_{6}$, BF${}_{4}$, ReO${}_{4}$, and SCN)}},
  author = {Salameh, B. and Yasin, S. and Dumm, M. and Untereiner, G. and Montgomery, L. and Dressel, M.},
  journal = {Phys. Rev. B},
  volume = {83},
  issue = {20},
  pages = {205126},
  numpages = {8},
  year = {2011},
  month = {May},
  publisher = {American Physical Society},
  doi = {10.1103/PhysRevB.83.205126},
  url = {https://link.aps.org/doi/10.1103/PhysRevB.83.205126}
}

@article{2022_Mugiraneza_chi_tutorial,
  title={Tutorial: a beginner’s guide to interpreting magnetic susceptibility data with the {Curie-Weiss} law},
  author={Mugiraneza, Sam and Hallas, Alannah M},
  journal={Communications Physics},
  volume={5},
  number={1},
  pages={95},
  year={2022},
  publisher={Nature Publishing Group UK London}
}

@article{2013_Qu_Class_Xover_PRB,
  title = {Structure and magnetism of {Cr$_{2}$[BP$_3$O$_{12}$]}: Towards the quantum-classical crossover in a spin-$\frac{3}{2}$ alternating chain},
  author = {Janson, O. and Chen, S. and Tsirlin, A. A. and Hoffmann, S. and Sichelschmidt, J. and Huang, Q. and Zhang, Z.J. and Tang, M.B. and Zhao, J.T. and Kniep, R. and Rosner, H.},
  journal = {Phys. Rev. B},
  volume = {87},
  issue = {6},
  pages = {064417},
  numpages = {11},
  year = {2013},
  month = {Feb},
  publisher = {American Physical Society},
  doi = {10.1103/PhysRevB.87.064417},
  url = {https://link.aps.org/doi/10.1103/PhysRevB.87.064417}
}

@article{1979_Exchange-coupled_pair_PRB,
  title = {Exchange-coupled pair model for the random-exchange {Heisenberg} antiferromagnetic chain},
  author = {Clark, W. G. and Tippie, L. C.},
  journal = {Phys. Rev. B},
  volume = {20},
  issue = {7},
  pages = {2914--2923},
  numpages = {0},
  year = {1979},
  month = {Oct},
  publisher = {American Physical Society},
  doi = {10.1103/PhysRevB.20.2914},
  url = {https://doi.org/10.1103/PhysRevB.20.2914}
}

@article{1995_Chi_Fit_JMMM,
title = {Finite segments in quasi-{1D} {H}eisenberg antiferromagnets: 
comparison of the isostructural systems {AgVP$_2$S$_6$ (S = 1)} and {AgCrP$_2$S$_6$ (S = 3/2)}},
journal = {Journal of Magnetism and Magnetic Materials},
volume = {140-144},
pages = {1677-1678},
year = {1995},
issn = {0304-8853},
doi = {10.1016/0304-8853(94)00524-9},
url = {https://doi.org/10.1016/0304-8853(94)00524-9},
author = {H. Mutka and C. Payen and P. Molini\'{e}}
}

@article{1997_Phase_Trans_Dimerized_spin_chains_Yamamoto_PRB,
  title = {Phase transitions in antiferromagnetic quantum spin chains with bond alternation},
  author = {Yamamoto, Shoji},
  journal = {Phys. Rev. B},
  volume = {55},
  issue = {6},
  pages = {3603--3612},
  numpages = {0},
  year = {1997},
  month = {Feb},
  publisher = {American Physical Society},
  doi = {10.1103/PhysRevB.55.3603},
  url = {https://link.aps.org/doi/10.1103/PhysRevB.55.3603}
}

@article{1999_S-half_dimerized_Tennant_PRB,
  title = {{$S=\frac{1}{2}$} alternating chain using multiprecision methods},
  author = {Barnes, T. and Riera, J. and Tennant, D. A.},
  journal = {Phys. Rev. B},
  volume = {59},
  issue = {17},
  pages = {11384--11397},
  numpages = {0},
  year = {1999},
  month = {May},
  publisher = {American Physical Society},
  doi = {10.1103/PhysRevB.59.11384},
  url = {https://link.aps.org/doi/10.1103/PhysRevB.59.11384}
}

@article{1996_S_3-2_Dimerized,
author = {Yajima ,Makoto and Takahashi ,Minoru},
title = { S= 3/2 Antiferromagnetic Spin Chain with Bond Alternation},
journal = {Journal of the Physical Society of Japan},
volume = {65},
number = {1},
pages = {39-42},
year = {1996},
doi = {10.1143/JPSJ.65.39},
URL = {https://doi.org/10.1143/JPSJ.65.39}
}

@article{1996_S1_Dimerization_Anis_Phase_Diag_JPSJ,
author = {Tonegawa ,Takashi and Nakao ,Takeshi and Kaburagi ,Makoto},
title = {Ground-State Phase Diagram and Magnetization Curves of the Spin-1 Antiferromagnetic {Heisenberg} Chain  with Bond Alternation and Uniaxial Single-Ion-Type Anisotropy},
journal = {Journal of the Physical Society of Japan},
volume = {65},
number = {10},
pages = {3317-3330},
year = {1996},
doi = {10.1143/JPSJ.65.3317},
URL = {https://doi.org/10.1143/JPSJ.65.3317}
}

@article{1994_S1_Dimerized_JPSJ,
author = {Kato ,Yusuke and Tanaka ,Akihiro},
title = {Numerical Study of the {S=1} Antiferromagnetic Spin Chain with Bond Alternation},
journal = {Journal of the Physical Society of Japan},
volume = {63},
number = {4},
pages = {1277-1280},
year = {1994},
doi = {10.1143/JPSJ.63.1277},
URL = {https://doi.org/10.1143/JPSJ.63.1277}
}

@article{2018_Single_Chain_TMT_Sci,
author = {Thang Pham  and Sehoon Oh  and Patrick Stetz  and Seita Onishi  and Christian Kisielowski  and Marvin L. Cohen  and Alex Zettl },
title = {Torsional instability in the single-chain limit of a transition metal trichalcogenide},
journal = {Science},
volume = {361},
number = {6399},
pages = {263-266},
year = {2018},
doi = {10.1126/science.aat4749},
URL = {https://www.science.org/doi/abs/10.1126/science.aat4749}
}

@article{2021_NbTe3_VTe3_TiTe3_in_CNT_JACS,
	title = {Stabilization of {NbTe$_3$}, {VTe$_3$}, and {TiTe$_3$} via Nanotube Encapsulation},
	volume = {143},
	issn = {0002-7863},
	url = {https://doi.org/10.1021/jacs.0c10175},
	doi = {10.1021/jacs.0c10175},
	number = {12},
	journal = {Journal of the American Chemical Society},
	author = {Stonemeyer, Scott and Cain, Jeffrey D. and Oh, Sehoon and Azizi, Amin and Elasha, Malik and Thiel, Markus and Song, Chengyu and Ercius, Peter and Cohen, Marvin L. and Zettl, Alex},
	month = mar,
	year = {2021},
	note = {Publisher: American Chemical Society},
	pages = {4563--4568},
}

@article{2025_Tunable_Synthesis_VxTey_Magnets_CNTs_NCom,
	title = {Tunable synthesis of atomic one-dimensional {V$_x$Te$_y$} magnets within single-walled carbon nanotubes},
	volume = {16},
	issn = {2041-1723},
	url = {https://doi.org/10.1038/s41467-025-61591-7},
	doi = {10.1038/s41467-025-61591-7},
	number = {1},
	journal = {Nature Communications},
	author = {Lan, Xuhua and Geng, Lin and Zhang, Zhen and Li, Yunfei and Yuan, Jian and Zhou, Chen-Xu and Huang, Song and Hu, Ziyi and Li, Jing and Yang, Chengpeng and Zhang, Yong and Fan, Zhaochuan and Tian, Dan and Zhao, Xiaoxu and Li, Qingwen and Kang, Lixing},
	month = jul,
	year = {2025},
	pages = {6300},
}

@article{2025_High-Spin_State_CrX2_JACS,
author = {Lee, Yangjin and Li, Linxuan and Zhang, Weihan and Choi, Uje and Lee, Kihyun and Kim, Young-Min and Ji, Wei and Zhou, Wu and Kim, Kwanpyo and Zettl, Alex},
title = {Robust High-Spin State in One-Dimensional {CrX$_2$} ({X = Cl, Br, I}) at the Single-Chain Limit},
journal = {Journal of the American Chemical Society},
volume = {147},
number = {30},
pages = {26776-26785},
year = {2025},
doi = {10.1021/jacs.5c07656},
note ={PMID: 40663099},
URL = {https://doi.org/10.1021/jacs.5c07656}, 
eprint = {https://doi.org/10.1021/jacs.5c07656}
}

@article{2019AgVOAsO4,
  title = {Field-induced double dome and Bose-Einstein condensation in the crossing quantum spin chain system {${\mathrm{AgVOAsO}}_{4}$}},
  author = {Weickert, Franziska and Aczel, Adam A. and Stone, Matthew B. and Garlea, V. Ovidiu and Dong, Chao and Kohama, Yoshimitsu and Movshovich, Roman and Demuer, Albin and Harrison, Neil and Gam\ifmmode \dot{z}\else \.{z}\fi{}a, Monika B. and Steppke, Alexander and Brando, Manuel and Rosner, Helge and Tsirlin, Alexander A.},
  journal = {Phys. Rev. B},
  volume = {100},
  issue = {10},
  pages = {104422},
  numpages = {8},
  year = {2019},
  month = {Sep},
  publisher = {American Physical Society},
  doi = {10.1103/PhysRevB.100.104422},
  url = {https://link.aps.org/doi/10.1103/PhysRevB.100.104422}
}

@article{2000_coppernitrate,
  title = {Triplet Waves in a Quantum Spin Liquid},
  author = {Xu, Guangyong and Broholm, C. and Reich, Daniel H. and Adams, M. A.},
  journal = {Phys. Rev. Lett.},
  volume = {84},
  issue = {19},
  pages = {4465--4468},
  numpages = {0},
  year = {2000},
  month = {May},
  publisher = {American Physical Society},
  doi = {10.1103/PhysRevLett.84.4465},
  url = {https://link.aps.org/doi/10.1103/PhysRevLett.84.4465}
}

@article{2022_1D_Review_Mat_Today,
title = {One-dimensional van der {Waals} quantum materials},
journal = {Materials Today},
volume = {55},
pages = {74-91},
year = {2022},
issn = {1369-7021},
doi = {10.1016/j.mattod.2022.03.015},
url = {https://doi.org/10.1016/j.mattod.2022.03.015},
author = {Alexander A. Balandin and Fariborz Kargar and Tina T. Salguero and Roger K. Lake}
}

@article{2024_1D_vdW_AAB_AdvMat,
  title={Achieving the {1D} Atomic Chain Limit in van  der {Waals} Crystals},
  author={Teeter, Jordan and Kim, Na Yeon and Debnath, Topojit and Sesing, Nicholas and Geremew, Tekwam and Wright, Dylan and Chi, Miaofang and Stieg, Adam Z and Miao, Jianwei and Lake, Roger K and Salguero, Tina and Balandin, Alexander A.},
  journal={Advanced Materials},
  volume={36},
  number={48},
  pages={2409898},
  year={2024},
  publisher={Wiley Online Library}
}

@article{1986_ZrSe3andZrS3,
    author = {R. Provencher and C. Ayache and S. Jandl and J.-P. Jay-Gerin},
    title = {Low-temperature specific heats of {ZrSe$_3$} and {ZrS$_3$}},
    journal = {Solid State Communications},
    volume = {59},
    number = {8},
    pages = {553-556},
    year = {1986}
}

@article{2022_XiChen_ZrSe3,
    author = {Youmind Xu and Shucheng Guo and Xi Chen},
    title = {Crystal Growth and Thermal Properties of Quasi-One-Dimensional van der {Waals} Material {ZrSe$_3$}},
    journal = {Micromachines},
    volume = {13},
    number = {11},
    pages = {1994},
    year = {2022}
}

@article{Bain2008,
  author  = {Bain, Gordon A. and Berry, John F.},
  title   = {Diamagnetic Corrections and {P}ascal's Constants},
  journal = {Journal of Chemical Education},
  volume  = {85},
  number  = {4},
  pages   = {532},
  year    = {2008},
  doi     = {10.1021/ed085p532}
}

@article{bethe1931,
  author = {Hans Bethe},
  title = {{Zur Theorie der Metalle. Eigenwerte und Eigenfunktionen der linearen Atomkette}},
  journal = {Zeitschrift für Physik},
  year = {1931},
  volume = {71},
  pages = {205--226}
}

@article{lieb1961,
  author = {Elliott H. Lieb and Theodore Schultz and Daniel C. Mattis},
  title = {Two soluble models of an antiferromagnetic chain},
  journal = {Annals of Physics},
  year = {1961},
  volume = {16},
  pages = {407--466}
}

@article{haldane1983,
  author = {F. Duncan M. Haldane},
  title = {Continuum dynamics of the {1-D} {Heisenberg} antiferromagnetic: {I}dentification with the {O(3)} nonlinear sigma model},
  journal = {Physics Letters A},
  year = {1983},
  volume = {93},
  pages = {464--468}
}

@article{2024_MoI3_Spin_Chains_JMMM,
title = {Magnetic spirals and biquadratic exchange in {1D MoX$_3$} spin chains},
journal = {Journal of Magnetism and Magnetic Materials},
volume = {594},
pages = {171882},
year = {2024},
issn = {0304-8853},
doi = {10.1016/j.jmmm.2024.171882},
url = {https://doi.org/10.1016/j.jmmm.2024.171882},
author = {Andy Mella and Eric Suárez-Morell and Alvaro S. Nunez}
}

@article{blochl1994projector,
  title = {Projector augmented-wave method},
  author = {Bl\"ochl, P. E.},
  journal = {Phys. Rev. B},
  volume = {50},
  issue = {24},
  pages = {17953--17979},
  numpages = {0},
  year = {1994},
  month = {Dec},
  publisher = {American Physical Society},
  doi = {10.1103/PhysRevB.50.17953},
  url = {https://link.aps.org/doi/10.1103/PhysRevB.50.17953}
}

@article{kresse1999ultrasoft,
  title = {From ultrasoft pseudopotentials to the projector augmented-wave method},
  author = {Kresse, G. and Joubert, D.},
  journal = {Phys. Rev. B},
  volume = {59},
  issue = {3},
  pages = {1758--1775},
  numpages = {0},
  year = {1999},
  month = {Jan},
  publisher = {American Physical Society},
  doi = {10.1103/PhysRevB.59.1758},
  url = {https://link.aps.org/doi/10.1103/PhysRevB.59.1758}
}

@article{kresse1996efficient,
  title = {Efficient iterative schemes for ab initio total-energy calculations using a plane-wave basis set},
  author = {Kresse, G. and Furthm\"uller, J.},
  journal = {Phys. Rev. B},
  volume = {54},
  issue = {16},
  pages = {11169--11186},
  numpages = {0},
  year = {1996},
  month = {Oct},
  publisher = {American Physical Society},
  doi = {10.1103/PhysRevB.54.11169},
  url = {https://link.aps.org/doi/10.1103/PhysRevB.54.11169}
}

@article{kresse1993ab,
  title = {Ab initio molecular dynamics for liquid metals},
  author = {Kresse, G. and Hafner, J.},
  journal = {Phys. Rev. B},
  volume = {47},
  issue = {1},
  pages = {558--561},
  numpages = {0},
  year = {1993},
  month = {Jan},
  publisher = {American Physical Society},
  doi = {10.1103/PhysRevB.47.558},
  url = {https://link.aps.org/doi/10.1103/PhysRevB.47.558}
}

@article{perdew1996generalized,
  title = {Generalized Gradient Approximation Made Simple},
  author = {Perdew, John P. and Burke, Kieron and Ernzerhof, Matthias},
  journal = {Phys. Rev. Lett.},
  volume = {77},
  issue = {18},
  pages = {3865--3868},
  numpages = {0},
  year = {1996},
  month = {Oct},
  publisher = {American Physical Society},
  doi = {10.1103/PhysRevLett.77.3865},
  url = {https://link.aps.org/doi/10.1103/PhysRevLett.77.3865}
}

@article{monkhorst1976special,
  title = {Special points for {Brillouin-zone} integrations},
  author = {Monkhorst, Hendrik J. and Pack, James D.},
  journal = {Phys. Rev. B},
  volume = {13},
  issue = {12},
  pages = {5188--5192},
  numpages = {0},
  year = {1976},
  month = {Jun},
  publisher = {American Physical Society},
  doi = {10.1103/PhysRevB.13.5188},
  url = {https://link.aps.org/doi/10.1103/PhysRevB.13.5188}
}

@article{grimme2010consistent,
author = {Grimme,Stefan  and Antony,Jens  and Ehrlich,Stephan  and Krieg,Helge },
title = {A consistent and accurate ab initio parametrization of density functional dispersion correction {(DFT-D)} for the 94 elements {H-Pu}},
journal = {The Journal of Chemical Physics},
volume = {132},
number = {15},
pages = {154104},
year = {2010},
doi = {10.1063/1.3382344},
URL = {https://doi.org/10.1063/1.3382344}
}

@ARTICLE{Phonopy,
   author       = "A. Togo and F. Oba and I. Tanaka", 
   title        = {First-principles calculations of the ferroelastic transition between rutile-type and {CaCl$_2$}-type {SiO$_2$} at high pressures}, 
   journal      = "Phys. Rev. B", 
   volume       = "78", 
   pages        = "134106", 
   year         = "2008", 
}

@misc{zheludev2007reslib,
  author       = {Andrey Zheludev},
  title        = {ResLib: Resolution calculation library for neutron scattering experiments},
  year         = {2007},
  version      = {2007},
  howpublished = {Computer software}
}

@inproceedings{ACCESS,
author = {Boerner, Timothy J. and Deems, Stephen and Furlani, Thomas R. and Knuth, Shelley L. and Towns, John},
title = {{ACCESS}: Advancing Innovation: {NSF's} Advanced Cyberinfrastructure Coordination Ecosystem: Services \& Support},
year = {2023},
isbn = {9781450399852},
publisher = {Association for Computing Machinery},
address = {New York, NY, USA},
url = {https://doi.org/10.1145/3569951.3597559},
doi = {10.1145/3569951.3597559},
booktitle = {Practice and Experience in Advanced Research Computing},
pages = {173-176},
numpages = {4},
location = {Portland, OR, USA},
series = {PEARC '23}
}

@article{2025_Qu_Sp_Dyn_1D_vdW_JACS,
author = {Li, Jing and Zhang, Zhen and Li, Yunfei and Zhang, Hui and Zhong, Yunlei and Li, Alei and Teng, Yu and Yao, Jian and Zhou, Chao and Fan, Zhaochuan and Geng, Lin and Kang, Lixing},
title = {Quantum Spin Dynamics of One-Dimensional Magnetic van der {Waals} Heterostructures},
journal = {Journal of the American Chemical Society},
volume = {147},
number = {27},
pages = {23972-23979},
year = {2025},
doi = {10.1021/jacs.5c07286},
note ={PMID: 40571493},
URL = {https://doi.org/10.1021/jacs.5c07286}
}

@article{Single-chain_encap,
author = {Milligan, Griffin M. and Singsen, Sirisak and To, Sydney and Aoki, Toshihiro and Zhi, Brian Y. and Collins, Cameron J. and Ogura, Kaleolani S. and Lee, Elizabeth M. Y. and Arguilla, Maxx Q.},
title = {Precision Synthesis of a Single Chain Polymorph of a {2D} Solid within Single-Walled Carbon Nanotubes},
journal = {Advanced Materials},
volume = {37},
number = {28},
pages = {2502213},
doi = {https://doi.org/10.1002/adma.202502213},
url = {https://advanced.onlinelibrary.wiley.com/doi/abs/10.1002/adma.202502213},
year = {2025}
}

@article{Random_exchange_Haldane,
  title = {Random-exchange Heisenberg behavior in the electron-doped quasi-one-dimensional spin-1 chain compound \text{AgVP$_2$S$_6$}},
  author = {Orban, Peter T. and Bernier, Shannon M. and Berry, Tanya and Siegler, Maxime A. and McQueen, Tyrel M.},
  journal = {Phys. Rev. B},
  volume = {110},
  issue = {5},
  pages = {054423},
  numpages = {10},
  year = {2024},
  month = {Aug},
  publisher = {American Physical Society},
  doi = {10.1103/PhysRevB.110.054423},
  url = {https://link.aps.org/doi/10.1103/PhysRevB.110.054423}
}

@article{Dudarev,
  title = {Electron-energy-loss spectra and the structural stability of nickel oxide:  An {LSDA+U} study},
  author = {Dudarev, S. L. and Botton, G. A. and Savrasov, S. Y. and Humphreys, C. J. and Sutton, A. P.},
  journal = {Phys. Rev. B},
  volume = {57},
  issue = {3},
  pages = {1505--1509},
  numpages = {0},
  year = {1998},
  month = {Jan},
  publisher = {American Physical Society},
  doi = {10.1103/PhysRevB.57.1505},
  url = {https://link.aps.org/doi/10.1103/PhysRevB.57.1505}
}

@misc{QD_MPMS_Notes,
note = {{\mbox Quantum Design, MPMS Application Note 1014-210: Oxygen Contamination, (1997)}}}

@misc{SM,
note = {{\mbox See Supplemental Material at http://URL for additional figures and data related to sample synthesis, susceptibility, XRD, XPS, binding energies, 
elastic neutron scattering peaks, DMRG calculations, sample
preparation, and oxygen contamination.}}}

@article{gapTdep,
  title = {Temperature-dependent spin gap and singlet ground state in {${\mathrm{BaCuSi}}_{2}$${\mathrm{O}}_{6}$}},
  author = {Sasago, Y. and Uchinokura, K. and Zheludev, A. and Shirane, G.},
  journal = {Phys. Rev. B},
  volume = {55},
  issue = {13},
  pages = {8357--8360},
  numpages = {0},
  year = {1997},
  month = {Apr},
  publisher = {American Physical Society},
  doi = {10.1103/PhysRevB.55.8357},
  url = {https://link.aps.org/doi/10.1103/PhysRevB.55.8357}
}

@article{SEQUOIA,
    doi = {10.1088/1742-6596/251/1/012058},
    url = {https://doi.org/10.1088/1742-6596/251/1/012058},
    year = {2010},
    month = {nov},
    publisher = {},
    volume = {251},
    number = {1},
    pages = {012058},
    author = {Granroth, G E and Kolesnikov, A I and Sherline, T E and Clancy, J P and Ross, K A and Ruff, J. P. C. and Gaulin, B. D. and Nagler, S. E.},
    title = {SEQUOIA: A Newly Operating Chopper Spectrometer at the {SNS}},
    journal = {Journal of Physics: Conference Series},
}

@article{Stone2025,
  author       = {Stone, Matthew B. and Granroth, Garrett E. and Pajerowski, Daniel M. and Abernathy, Douglas L. and Conner, David L. and DeBeer-Schmitt, Lisa and Fanelli, Victor R. and Goyette, Richard and Kolesnikov, Alexander I. and Mills, Rebecca and Odom, Mary and Podlesnyak, Andrey and Schmitt, Christopher and Sherline, Todd E. and Solomon, Landon and Wenzel, John F.},
  title        = {Sample changers for direct geometry neutron chopper spectrometers},
  journal      = {Scientific Reports},
  year         = {2025},
  volume       = {15},
  number       = {1},
  pages        = {31936},
  doi          = {10.1038/s41598-025-17049-3},
  url          = {https://doi.org/10.1038/s41598-025-17049-3},
  issn         = {2045-2322},
  date         = {2025-08-29}
}

@article{HAN2024109288,
title = {{INSPIRED}: Inelastic neutron scattering prediction for instantaneous results and experimental design},
journal = {Computer Physics Communications},
volume = {304},
pages = {109288},
year = {2024},
issn = {0010-4655},
doi = {https://doi.org/10.1016/j.cpc.2024.109288},
url = {https://www.sciencedirect.com/science/article/pii/S001046552400211X},
author = {Bowen Han and Andrei T. Savici and Mingda Li and Yongqiang Cheng}
}

@article{yoon2020jx,
  title={Jx: An open-source software for calculating magnetic interactions based on magnetic force theory},
  author={Yoon, Hongkee and Kim, Taek Jung and Sim, Jae-Hoon and Han, Myung Joon},
  journal={Computer Physics Communications},
  volume={247},
  pages={106927},
  year={2020},
  publisher={Elsevier},
  doi = {10.1016/j.cpc.2019.106927}
}

@article{liechtenstein1987local,
  title={Local spin density functional approach to the theory of exchange interactions in ferromagnetic metals and alloys},
  author={Liechtenstein, A Il and Katsnelson, MI and Antropov, VP and Gubanov, VA},
  journal={Journal of Magnetism and Magnetic Materials},
  volume={67},
  number={1},
  pages={65--74},
  year={1987},
  publisher={Elsevier}
}

@article{pizzi2020wannier90,
  title={Wannier90 as a community code: new features and applications},
  author={Pizzi, Giovanni and Vitale, Valerio and Arita, Ryotaro and Bl{\"u}gel, Stefan and Freimuth, Frank and G{\'e}ranton, Guillaume and Gibertini, Marco and Gresch, Dominik and Johnson, Charles and Koretsune, Takashi and Ibañez-Azpiroz, Julen and Lee, Hyungjun and Lihm, Jae-Mo and Marchand, Daniel and Marrazzo, Antimo and Mokrousov, Yuriy and Mustafa, Jamal I and Nohara, Yoshiro and Nomura, Yusuke and Paulatto, Lorenzo and Poncé, Samuel and Ponweiser, Thomas and Qiao, Junfeng and Thöle, Florian and Tsirkin, Stepan S and Wierzbowska, Małgorzata and Marzari, Nicola and Vanderbilt, David and Souza, Ivo and Mostofi, Arash A. and Yates, Jonathan R.},
  journal={Journal of Physics: Condensed Matter},
  volume={32},
  number={16},
  pages={165902},
  year={2020},
  publisher={IOP Publishing}
}

@article{togo2024spglib,
  title={Spglib: a software library for crystal symmetry search},
  author={Togo, Atsushi and Shinohara, Kohei and Tanaka, Isao},
  journal={Science and Technology of Advanced Materials: Methods},
  volume={4},
  number={1},
  pages={2384822},
  year={2024},
  publisher={Taylor \& Francis}
}

@article{1981_Mag_solid_O2,
  title = {Magnetism of solid oxygen},
  author = {DeFotis, Gary C.},
  journal = {Phys. Rev. B},
  volume = {23},
  issue = {9},
  pages = {4714--4740},
  numpages = {0},
  year = {1981},
  month = {May},
  publisher = {American Physical Society},
  doi = {10.1103/PhysRevB.23.4714},
  url = {https://link.aps.org/doi/10.1103/PhysRevB.23.4714}
}

@article{2017_H_T_Phase_diag_O2,
  title = {{$H\text{\ensuremath{-}}T$} phase diagram of solid oxygen},
  author = {Nomura, T. and Matsuda, Y. H. and Kobayashi, T. C.},
  journal = {Phys. Rev. B},
  volume = {96},
  issue = {5},
  pages = {054439},
  numpages = {5},
  year = {2017},
  month = {Aug},
  publisher = {American Physical Society},
  doi = {10.1103/PhysRevB.96.054439},
  url = {https://link.aps.org/doi/10.1103/PhysRevB.96.054439}
}

\end{document}